\documentclass[11pt,dvips]{article}
\usepackage{graphicx}
\usepackage{amsmath}
\usepackage{amssymb}
\usepackage{latexsym}
\usepackage{color}
\begin{document}
\thispagestyle{empty}
\begin{center}

\vspace{1.8cm}

 {\large {\bf Quantum discord in  photon added
  Glauber coherent states of ${\rm GHZ}$-type}}\\

\vspace{1.5cm}

{\bf  M. Daoud}$^{a,b,c}${\footnote { email: {\sf m$_-$daoud@hotmail.com}}}, {\bf
W. Kaydi}$^{d,e}$ {\footnote { email: {\sf kaydi.smp@gmail.com}}} and {\bf
     H. El Hadfi }$^{d,e}$ {\footnote { email: {\sf hanane.elhadfi@gmail.com}}}

\vspace{0.5cm}
$^{a}${\it Max Planck Institute for the Physics of Complex Systems, Dresden, Germany}\\[0.5em]
$^{b}${\it Abdus Salam  International Centre for Theoretical Physics, Miramare, Trieste, Italy}\\[0.5em]
$^{c}${\it Department of Physics , Faculty of Sciences, University Ibnou Zohr,
 Agadir,
Morocco}\\[0.5em]
$^{d}${\it LPHE-Modeling and Simulation, Faculty  of Sciences,
University
Mohammed V, Rabat, Morocco}\\[0.5em]
$^{e}${\it Centre of Physics and Mathematics (CPM),
University
Mohammed V, Rabat, Morocco}\\[0.5em]

\vspace{3cm} {\bf Abstract}
\end{center}
\baselineskip=18pt
\medskip

We investigate the influence of photon excitations on quantum
correlations in tripartite Glauber coherent states of
Greenberger-Horne-Zeilinger type. The pairwise correlations are
measured by means of the entropy-based  quantum discord. We also
analyze the monogamy property of quantum discord in this class of
tripartite states in terms of the strength of Glauber coherent
states and the photon excitation order.

\newpage

\section{Introduction}

In the context of information processing and transmission, several theoretical and experimental
results confirm the advantages of quantum protocols compared to  their classical counterparts (see for instance \cite{Nielsen,Mermin,Wilde}).
Quantum technology exploiting  the intriguing  phenomena of quantum world, such as entanglement, offers secure ways
for communication \cite{Bennett1,Bennett2} and potentially  powerful algorithms in quantum computation \cite{Shor}. Originally, quantum information processing
focused on discrete (finite-dimensional) entangled states like the polarizations of a photon or
discrete levels of an atom. But, the extension
from discrete to continuous variables  has been also proven beneficial  in coding and manipulating efficiently quantum information.
Coherent states, which constitute the prototypical instance of  continuous-variables  states, are
expected to play a central role in this context. They are appealing for
their mathematical elegance (continuity and over-completion property)
and closeness to classical  physical states (minimization of Heisenberg uncertainty relation).  Implementing
a logical qubit encoding by treating entangled
coherent states  as  qubits in a two dimensional Hilbert space has been shown a promising strategy
in performing successfully  various quantum tasks such as  quantum  teleportation
\cite{Enk,JKL01},
 quantum computation \cite{JK,Ralph,Ralph2},
entanglement purification \cite{JKpuri} and errors correction
\cite{Glancy}.  In view of these potential applications, a special
attention was paid, during the last years,  to the identification,
characterization and quantification of  quantum correlations in
bipartite coherent states systems (see for instance the papers
\cite{Sanders3,Sanders2,Wang} and references therein). The bipartite
treatment was extended to superpositions of multimode coherent
states \cite{Jex,Zheng,Wang1,daoud1,daoud2} which exhibit
multipartite entanglement.
One may quote for instance entanglement properties in {\rm GHZ}
(Greenberger-Horne-Zeilinger), {\rm W} (Werner) states discussed in
\cite{Jeong1,Li} and entangled coherent state extensions of cluster
qubits investigated in \cite{Munhoz,Wang-WF,Becerra}. To quantify quantum
correlations beyond entanglement in coherent states systems,
measures such as quantum discord
\cite{Ollivier-PRL88-2001,Vedral-et-al} and its geometric variant
\cite{Dakic2010} were used. Explicit results were derived for
quantum discord \cite{Luo,Ali,Adesso,Shi1,Girolami,Shi2,Rachid1,Rachid2}
and geometric quantum discord \cite{Adesso1,Giorda,Yin,Rachid3} for
some special sets of coherent states.\\

On the other hand, decoherence is a crucial process to understand
the emergence of classicality in quantum systems. It describes the
inevitable degradation of quantum correlations due to experimental
and environmental noise. Various decoherence models were
investigated and in particular  the phenomenon of entanglement
sudden death  was considered in a number of distinct contexts (see
for instance \cite{YuEberly-2007} and reference therein). For
optical qubits based on coherent states,  the influence of the
environment,  is mainly due to energy loss or photon
 absorption. The photon loss or equivalently amplitude damping in a noisy
 environment can be modeled by assuming that some of field energy and
 information is lost after transmission through a beam splitter
 \cite{Rachid1,Wickert}. Interestingly, it has been shown that a
 beam spitting device with a coherent in the first input and a
 number state in the second input generates photon-added coherent
 states \cite{Dakna}. Henceforth, understanding the influence of photon excitations
  might be useful to develop the adequate strategies in improving the performance of noise reduction in quantum
 processing protocols involving coherent states. In this sense, some
 authors considered the  concurrence as measure of the entanglement  in
bipartite and tripartite photon added
  coherent states \cite{Li,Xu}.\\

In this work, we  derive the analytical expression of pairwise
quantum discord in a three modes system initially prepared in a
tripartite Glauber coherent state of ${\rm GHZ}$-type. In
particular, we shall consider the influence of photon excitation of
a single mode on the dynamics of pairwise quantum correlations.
Mathematically, this process  is represented by the action of a
suitable creation operator on the  state of the first subsystem.
Another important issue in photon added ${\rm GHZ}$-type coherent
states
 concerns  the distribution of quantum
discord between the different parts of
 the whole system.  In fact, we study  the shareability of quantum
correlations which  obeys a restrictive inequality termed  in the literature as the monogamy property \cite{Coffman} (see
also \cite{Giorgi,Prabhu,Sudha,Allegra,Ren,Bruss}).\\

This paper is organized as follows. In section 2, basic definitions and
equations related to photon added coherent states are presented.
We also consider the quantum correlations as measured by the entanglement of formation in quasi-Bell states.
In particular, we introduce  an encoding map  to pass from continuous
variables (coherent states) to  discrete variables (logical quantum bits).  Along the same
line of reasoning, this qubit encoding is extended, in section 3, to tripartite photon added coherent states of
${\rm GHZ}$-type. The pairwise quantum discord quantifying the amount of quantum correlations
existing in the system is analytically derived.
In section 4,  we study the monogamy property of quantum discord. Numerical
illustrations of the monogamy inequality  are presented in
 some special cases. Concluding remarks close this paper.

\section{Entanglement in photon added quasi-Bell states}

\subsection{Photon added coherent states and qubit mapping}

The basic objects in this work are the Glauber coherent states $\vert \alpha \rangle $ and
$\vert -\alpha \rangle $ where $\alpha$ is a complex number which determines the coherent
amplitude of the electromagnetic field. Mathematically, a  single-mode quantized radiation field
is represented by the harmonic oscillator algebra spanned by the creation $a^+$ and annihilation $a^-$
operators. The process of adding $m$ photons to coherent states of type $\vert \alpha \rangle $ and
$\vert -\alpha \rangle $  is usually represented by the action  of the operator $(a^+)^m$ ($m$ is a non negative integer) \cite{Agarwal}.
Several experimental as well theoretical studies were devoted to the generation and nonclassical
properties of photon-added coherent states \cite{Zavatta} (for a recent review see \cite{Kim}). Explicitly,
$m$ successive  actions of creation operator $a^+$  on  the  Glauber coherent states
\begin{equation}
|\alpha
\rangle=e^{-\frac{|\alpha|^2}{2}}\sum^{\infty}_{n=0}\frac{\alpha^{n}}{\sqrt{n!}}|n\rangle,
\end{equation}
lead to the un-normalized states
\begin{equation}\label{vecteur 1}
||\alpha,m\rangle=\left( a^{+}\right) ^{m}\mid \alpha \rangle = e^{-\frac{|\alpha|^2}{2}} \sum_{n=0}^{\infty }\frac{ \alpha^{n}}{n!}\sqrt{
\left( n+m\right) !}\mid n+m \rangle.
\end{equation}
 The vectors $\vert n \rangle$ denote the Fock-Hilbert states
 of the harmonic oscillator. The normalized $m$-photon added coherent states are defined by
\begin{equation}\label{vecteur 2}
|\alpha , m \rangle = \frac{(a^{+})^{m} \vert \alpha \rangle}
{\sqrt{\langle \alpha \vert (a^{-})^{m} (a^{+})^{m} \vert \alpha \rangle}}  ,
\end{equation}
where the quantity
\begin{equation}\label{valeur-moy}
\langle\alpha|(a^{-})^{m}(a^{+})^{m}|\alpha\rangle  =  m! L_{m}(-| \alpha|^{2} ),
\end{equation}
involves the Laguerre polynomial of order $m$ defined by
\begin{equation}\label{laguerre}
L_{m}(x)=\sum_{n=0}^{m}\frac{(-1)^{n} m! x^{n}}{(n!)^{2}(m-n)!}.
\end{equation}
Photon added coherent states interpolate between  electromagnetic field coherent states (quasi-classical states)
and Fock states $\vert n \rangle$ (purely quantum states). Furthermore, they exhibit non-classical features such
as squeezing, negativity of Wigner distribution and sub Poissonian statistics \cite{Kim}. Their experimental generation
using parametric down conversion in a nonlinear crystal was reported in \cite{Zavatta}.  Photon-coherent states
$|\alpha , m \rangle$ and $|-\alpha , m \rangle$, of the same amplitude and phases differing by $\pi$,
are not orthogonal to each other. Indeed using the expression
\begin{equation}\label{valeur-moy1}
\langle-\alpha|(a^{-})^{m}(a^{+})^{m}|\alpha\rangle  =  e^{-2| \alpha |^{2}} m! L_{m}(| \alpha|^{2}),
\end{equation}
it is simply verified that the overlap between the two states is
\begin{equation}\label{over-lap}
\langle - \alpha , m  |\alpha , m \rangle = e^{-2| \alpha |^{2}} \frac{L_{m}(| \alpha|^{2}) }{L_{m}(-| \alpha|^{2})}.
\end{equation}
Considering the nonorthogonality property (\ref{over-lap}), the
identification of photon added coherent states $|\alpha , m \rangle$
and $|-\alpha , m \rangle$ as basis of a logical qubit is only
possible for  $\vert \alpha \vert$ large $(\vert \alpha \vert \geq
2)$. Alternatively, the Schr\"odinger cat states, the even and odd
coherent states, can be used to encode a qubit.  Indeed, based on
the encoding scheme proposed in \cite{Ralph2}, we introduce a two
dimensional basis spanned by the orthogonal qubits $\vert + , m
\rangle $ and $\vert - , m \rangle $ defined by
\begin{equation}\label{qubit-m}
 \vert \pm , m \rangle = \frac{1}{\sqrt{2 \pm 2 \kappa_m e^{-2|\alpha|^{2}}}} ~ (\vert \alpha , m\rangle \pm \vert -\alpha , m\rangle)
\end{equation}
where
\begin{equation}\label{kappa}
\kappa_m \equiv \kappa_m (|\alpha|^{2}) :=  \frac{L_{m}(|\alpha|^{2})}{L_{m}(-|\alpha|^{2})}.
\end{equation}
Clearly, for $m = 0$, one has  $\kappa_0 = 1$ and  the logical qubits (\ref{qubit-m}) reduce to
\begin{equation}\label{qubit-0}
\vert \pm  \rangle= \frac{1}{\sqrt{2 \pm 2 e^{-2|\alpha|^{2}}}} ~ (\vert \alpha \rangle \pm \vert -\alpha \rangle),
\end{equation}
which coincide with even and odd Glauber coherent states providing the qubit encoding scheme introduced in \cite{Ralph2}.
This qubit encoding is important
 in dealing with quantum correlation in photon added coherent states
and to investigate the influence of the photon  excitations processes.  To illustrate  this, we shall first consider the
entanglement in quasi-Bell states which are very interesting in quantum optics and serve as valuable resource
 for quantum teleportation and many other quantum computing operations. The quasi-Bell states
\begin{equation}\label{Bell-0}
|{\rm B}_{k}(\alpha)\rangle={\cal N}_{k}(\alpha)~ \big[
|\alpha\rangle\otimes|\alpha\rangle + e^{ik\pi} |-\alpha\rangle\otimes|-\alpha\rangle\big],
\end{equation}
with $k = 0  ~({\rm mod}~2)$ (resp.  $k = 1 ~({\rm mod}~2)$) stands for even (resp. odd) quasi-Bell states and the
normalization factor ${\cal N}_{k}(\alpha)$ is
\begin{equation}\label{Norme-0}
{\cal N}^{-2}_{k}(\alpha) =  2 + 2  e^{-4\vert \alpha \vert^2} \cos k\pi.
\end{equation}
By  repeated actions of the creation operator on the first mode, the resulting
excited quasi-Bell states are
\begin{equation}
||{\rm B}_{k}(\alpha,m)\rangle = {\cal N}_{k}(\alpha)~  \bigg[ \big[(a^{+})^{m}\otimes \mathbb{I}\big]
|\alpha\rangle\otimes|\alpha\rangle + e^{ik\pi} ~ \big[(a^{+})^{m} \otimes \mathbb{I}\big]|-\alpha\rangle\otimes|-\alpha\rangle\bigg]
\end{equation}
are un-normalized ($\mathbb{I}$ stands for the unity operator). Using the norm of the vectors $||{\rm B}_{k}(\alpha,m)\rangle$ given by
\begin{equation}
\langle {\rm B}_{k}(\alpha,m) || {\rm B}_{k}(\alpha,m) \rangle = m!~ \frac{L_{m}(-|\alpha|^2) +  e^{-4|\alpha|^{2}}L_{m}(|\alpha|^2)~\cos k\pi}{1 +  e^{-4\vert \alpha \vert^2} \cos k\pi},
\end{equation}
 we introduce the normalized photon-added quasi-Bell states as
 \begin{equation}
|{\rm B}_{k}(\alpha,m)\rangle= \frac{|| {\rm B}_{k}(\alpha,m) \rangle }{\sqrt{\langle {\rm B}_{k}(\alpha,m) || {\rm B}_{k}(\alpha,m) \rangle }}.
\end{equation}
 They can be rewritten as
\begin{equation}\label{Bell-m}
|{\rm B}_{k}(\alpha,m)\rangle=\mathcal{N}_{k}(\alpha,m)~
\big[|m, \alpha \rangle \otimes|\alpha\rangle + e^{ik\pi} ~|m , -\alpha \rangle \otimes|-\alpha\rangle\big],
\end{equation}
 in terms of the normalized photon added coherent state (\ref{vecteur 2}). The normalization factor in (\ref{Bell-m}) is
\begin{equation}\label{Norme-m}
 \mathcal{N}^{-2}_{k}(\alpha,m) = 2 + 2 \kappa_m e^{-4\vert \alpha \vert^2} \cos k\pi,
\end{equation}
which reduces for  $m=0$  to (\ref{Norme-0}) and the quasi-Bell states (\ref{Bell-0}) are recovered.
\subsection {Dynamics of the entanglement of formation under photon excitation}
Using the qubit mapping (\ref{qubit-m}) for the first mode and (\ref{qubit-0}) for the second,
 the bipartite state (\ref{Bell-m})  is converted in the two qubit state
\begin{equation}\label{eq C}
|{\rm B}_{k}(\alpha,m)\rangle=\mathcal{N}_{k}(\alpha,m)\sum_{i=\pm}\sum_{j=\pm} C_{ij} ~\vert i, m\rangle \otimes \vert j \rangle,
\end{equation}
where the vectors $\vert i, m\rangle $ (resp. $\vert j\rangle $) are defined by (\ref{qubit-m}) (resp. (\ref{qubit-0}))
and the expansion coefficients are given by
$$C_{++} = c^+_{m}c^+(1+e^{ik\pi}),\quad
C_{-+} = c^+c^-_{m}(1-e^{ik\pi}),\quad
C_{+-} = c^+_{m}c^-(1-e^{ik\pi}),\quad
C_{--} = c^-c^-_{m}(1+e^{ik\pi}),$$
with
$$ c^{\pm}_m = \sqrt{\frac{1 \pm \kappa_m e^{-2|\alpha|^{2}}}{2}} \qquad c^{\pm} = \sqrt{\frac{1 \pm e^{-2|\alpha|^{2}}}{2}}. $$
In a pure bipartite system, the
quantum discord coincides with entanglement of formation (see for instance \cite{Luo,Ali,Adesso}). Thus, to discuss the
effect of the photon excitations of quasi-Bell states (\ref{Bell-m}), we quantify
the quantum correlations  by means of the entanglement of formation.
 We recall that for $\rho_{12}$ the density
matrix for a pair of qubits~$1$ and~$2$ which may be pure or mixed, the entanglement of formation
is defined by ~\cite{Wootters98}
\begin{equation}\label{def-EOF}
E(\rho_{12}) = H(\frac{1}{2} + \frac{1}{2} \sqrt{1 - \vert  C(\rho_{12})\vert^2}),
\end{equation}
where  $H(x) = -x\log_{2} x -(1-x)\log_{2} (1-x)$ is the binary
entropy function. The concurrence $C(\rho_{12})$ is given by
\begin{equation}\label{def-CONC}
C(\rho_{12})=\max \left\{ \lambda _1-\lambda _2-\lambda _3-\lambda
_4,0\right\}
\end{equation}
for~$\lambda_1\ge\lambda_2\ge\lambda_3\ge\lambda_4$ the square roots
of the eigenvalues of the "spin-flipped" density matrix $\varrho_{12}\equiv\rho_{12}(\sigma_y\otimes\sigma_y)\rho_{12}^{\star}(\sigma_y\otimes \sigma_y)$
where the star stands for complex conjugation
and $\sigma_y $ is the usual Pauli matrix. In the state (\ref{Bell-m}), it
easy to check that the concurrence (\ref{def-CONC}) gives
\begin{equation}\label{conci}
C_{12}=2\mathcal{N}^{2}_{k}(\alpha,m)|C_{++}C_{--}-C_{+-}C_{-+}|,
\end{equation}
which rewrites explicitly as
\begin{equation}\label{Cshow}
C_{12}=\frac{\sqrt{1-e^{-4|\alpha|^{2}}}
\sqrt{ 1 - \kappa^2_m e^{-4|\alpha|^{2}}}}
{1 + \kappa_m  e^{-4|\alpha|^{2}}\cos k\pi }
\end{equation}
in terms of the coherent states amplitude  $|\alpha|$ and the excitation order $m$. This
result coincides with one obtained in \cite{Xu} using a different qubit encoding.  It follows that
entanglement of formation is
\begin{equation}\label{EOF-Bell-m}
E_{12} ~= ~ H\Bigg[ \frac{1}{2} + \frac{e^{-2|\alpha|^2}(1 + \kappa_m\cos k\pi)}{2 + 2\kappa_me^{-4|\alpha|^2}\cos k\pi} \Bigg].
\end{equation}
For $m = 0$, one
has
\begin{equation}\label{Cshow}
C_{12} = \frac{ 1-e^{-4|\alpha|^{2}}}{1 + e^{-4|\alpha|^{2}}\cos k\pi}.
\end{equation}
To illustrate the influence of the photon excitation on the quantum correlation between
the modes of the quasi-Bell state (\ref{Bell-0}), we report in the figures 1 and 2 the
behavior of the entanglement of formation $E_{12}$ (\ref{EOF-Bell-m}) versus
 Glauber coherent states amplitude $|\alpha|^2$ and the overlap $p= \langle \alpha \vert -\alpha \rangle = e^{-2|\alpha|^2}$ for different values of $m$. We note
 that for $|\alpha|$ large $( |\alpha|^2 \geq 1.5)$, the entanglement of formation tends to unit independently of the
 number of added photons $m$. Indeed, from equation (\ref{EOF-Bell-m}), one gets $E_{12} = 1$ for $\vert \alpha \vert \longrightarrow \infty$.
 Note that, in this limit, the Glauber
 coherent states $\vert \alpha \rangle$ and $\vert -\alpha \rangle$ tends to
 orthogonality  and an orthogonal basis can be constructed such that $\vert {\bf 0} \rangle \equiv \vert \alpha \rangle $
 and $\vert {\bf 1} \rangle \equiv  \vert -\alpha \rangle $. Thus, in the strong regime $\vert \alpha \vert \longrightarrow \infty$,  the quasi-Bell states (\ref{Bell-0})
 become maximally entangled
$$ |{\rm B}_{k}(\infty)\rangle= \frac{1}{\sqrt{2}}~ \big[
| {\bf 0}  \rangle\otimes|{\bf 0}  \rangle + e^{ik\pi} |{\bf 1}  \rangle\otimes|{\bf 1} \rangle\big]. $$
Subsequently, maximally entangled quasi-Bell states are robust against any photon addition process. Another
interesting limiting situation concerns quasi-Bell states with smaller values of $\alpha$ (weak regime). For $\alpha \longrightarrow 0 $,
the symmetric ($k = 0~ ({\rm mod}~2)$)-quasi-Bell state  (\ref{Bell-0}) reduces to the separable state $\vert 0 \rangle \otimes \vert 0 \rangle  $
and by adding $m$ photons it becomes $\vert m \rangle \otimes \vert 0 \rangle$. No quantum correlation is created by the photon excitation $(E_{12} = 0)$. This
result can be also obtained from (\ref{EOF-Bell-m}) for $|\alpha| \longrightarrow 0$. As depicted in the figure 2, the
situation is completely different for anti-symmetric quasi-Bell states ($k = 1~ ({\rm mod}~2)$)  (\ref{Bell-0}). For $\alpha$ approaching zero, the
entanglement of formation decreases as the photon excitation number $m$ increases. For $|\alpha| \longrightarrow 0$, the Laguerre polynomial (\ref{laguerre})
can be approximated by $L_m(\vert \alpha \vert^2) \simeq 1 - m \vert \alpha \vert^2$ and the quantity (\ref{kappa}) writes
\begin{equation}\label{kappa-approx}
\kappa_m(\vert \alpha \vert^2) \simeq 1 - 2m \vert \alpha \vert^2.
\end{equation}
Reporting (\ref{kappa-approx}) in (\ref{EOF-Bell-m}), one gets
\begin{equation}\label{EOF-approx}
E_{12} ({\rm B}_{1}(0,m))  \simeq H\bigg(\frac{m+1}{m+2}\bigg).
\end{equation}
It is interesting to note that in the situation when $|\alpha| \longrightarrow 0$, the anti-symmetric quasi-Bell states  (\ref{Bell-0})
reduce to the maximally entangled two qubit state of ${\rm W}$-type
\begin{equation}\label{W12}
|{\rm B}_{1}(0)\rangle= \frac{1}{\sqrt{2}}~ \big[
| 0 \rangle\otimes| 1  \rangle +  |1   \rangle\otimes| 0 \rangle\big]
\end{equation}
where $\vert  0 \rangle$ and $\vert  1 \rangle$ denote the Fock number states. The action of
the  operator $(a^+)^m$  on the state $|{\rm B}_{1}(0)\rangle$  gives
$$ |{\rm B}_{1}(0,m)\rangle= \frac{1}{\sqrt{m+2}}~ \big[
| m \rangle\otimes| 1  \rangle +  \sqrt{m+1} |m+1   \rangle\otimes| 0 \rangle\big]. $$
In this case, the concurrence is
\begin{equation}\label{C12-approx}
C_{12} ({\rm B}_{1}(0,m)) = 2 ~\frac{\sqrt{m+1}}{m+2},
\end{equation}
which agrees with the result (\ref{EOF-approx}).  Clearly, adding photons to maximally entangled
states of ${\rm W}$-type (\ref{W12}) diminishes the amount of pairwise quantum correlations.
\noindent For the intermediate regime, corresponding to $|\alpha|^2$ ranging between 0 and 1.5, the entanglement of formation
increases as the Glauber coherent state amplitude $\alpha$ increases. We note that
adding photon process induces a quick activation of the creation of quantum correlation for the
symmetric quasi-Bell states (figure 1). Similarly,
the results presented in figure 2 show that
increasing the amplitude of  anti-symmetric quasi-Bell states tends to compensate the quantum
 correlation loss due to photon excitations in states of ${\rm W}$ type.
\begin{figure}[!ht]
\centering
\begin{minipage}[t]{3in}
\centering
\includegraphics[width=3in]{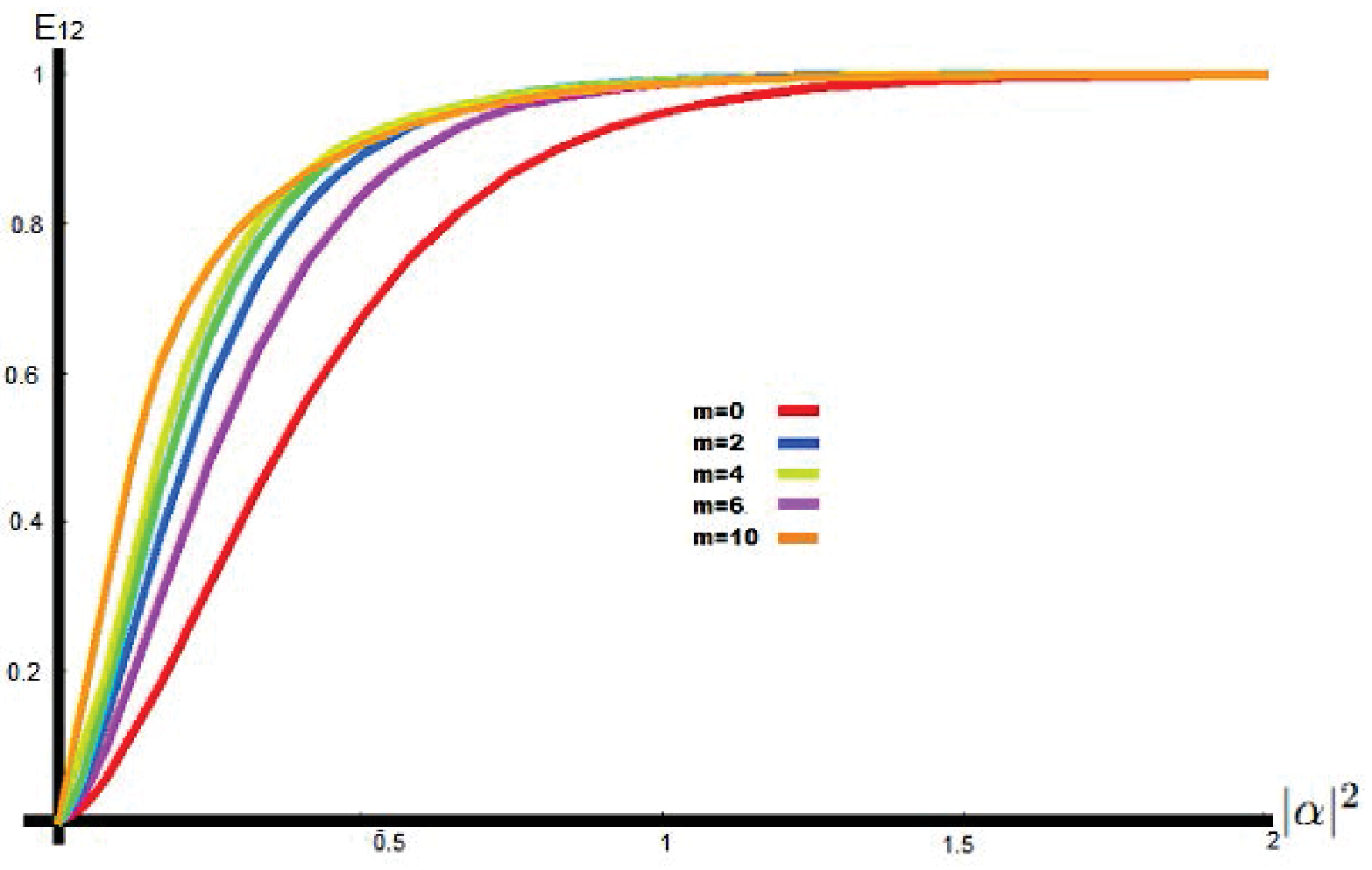}
\end{minipage}
\begin{minipage}[t]{3in}
\centering
 \includegraphics[width=3in]{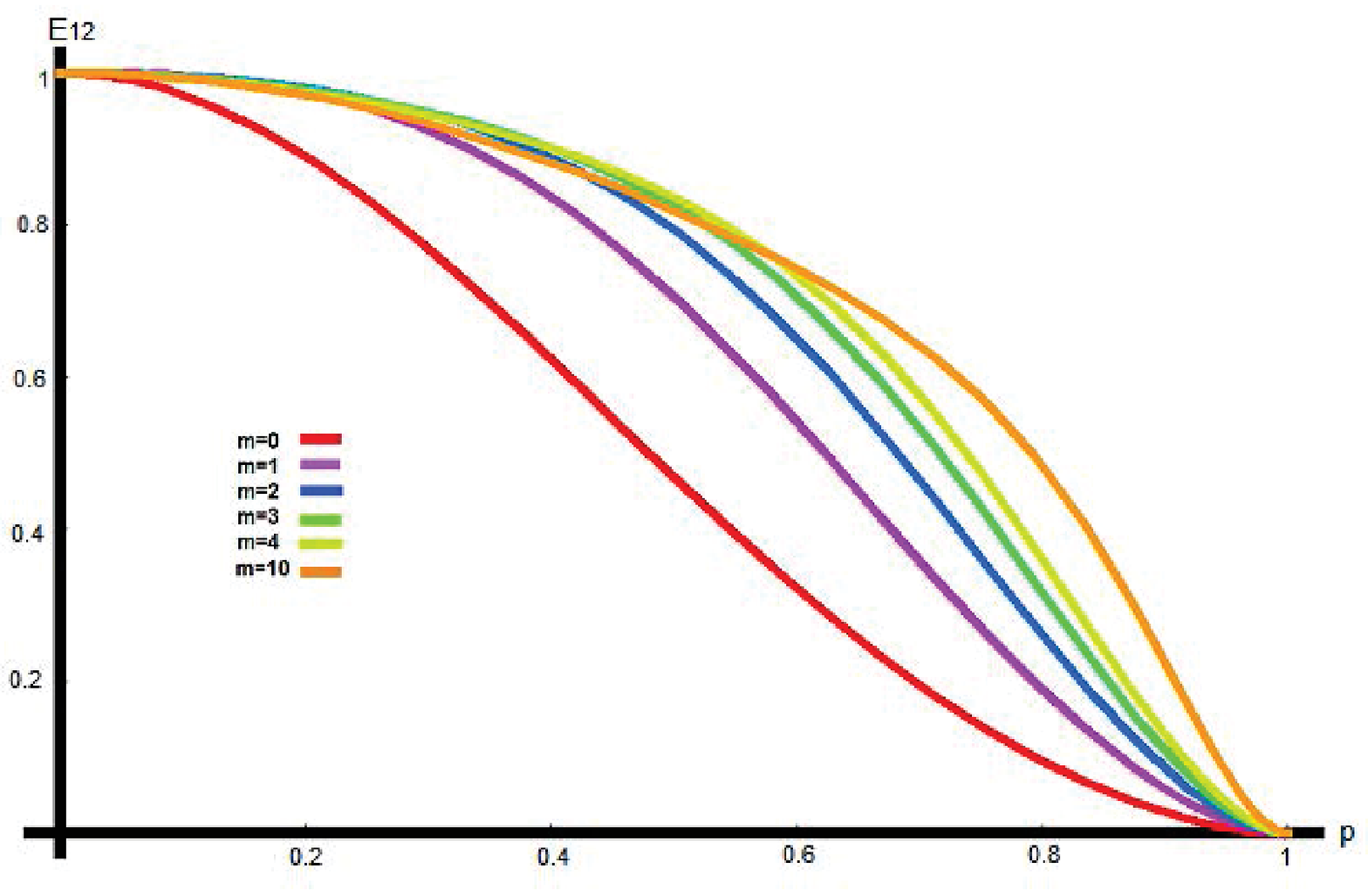}
\end{minipage}

{\bf Figure 1.}  {\sf The entanglement of formation $E_{12}$ versus
$|\alpha|^2$  and $p=e^{-2|\alpha|^2}$ for $k=0$ and different
values of photon excitation number $m$.}
\end{figure}
\begin{figure}[!ht]
\centering
\begin{minipage}[t]{3in}
\centering
\includegraphics[width=3in]{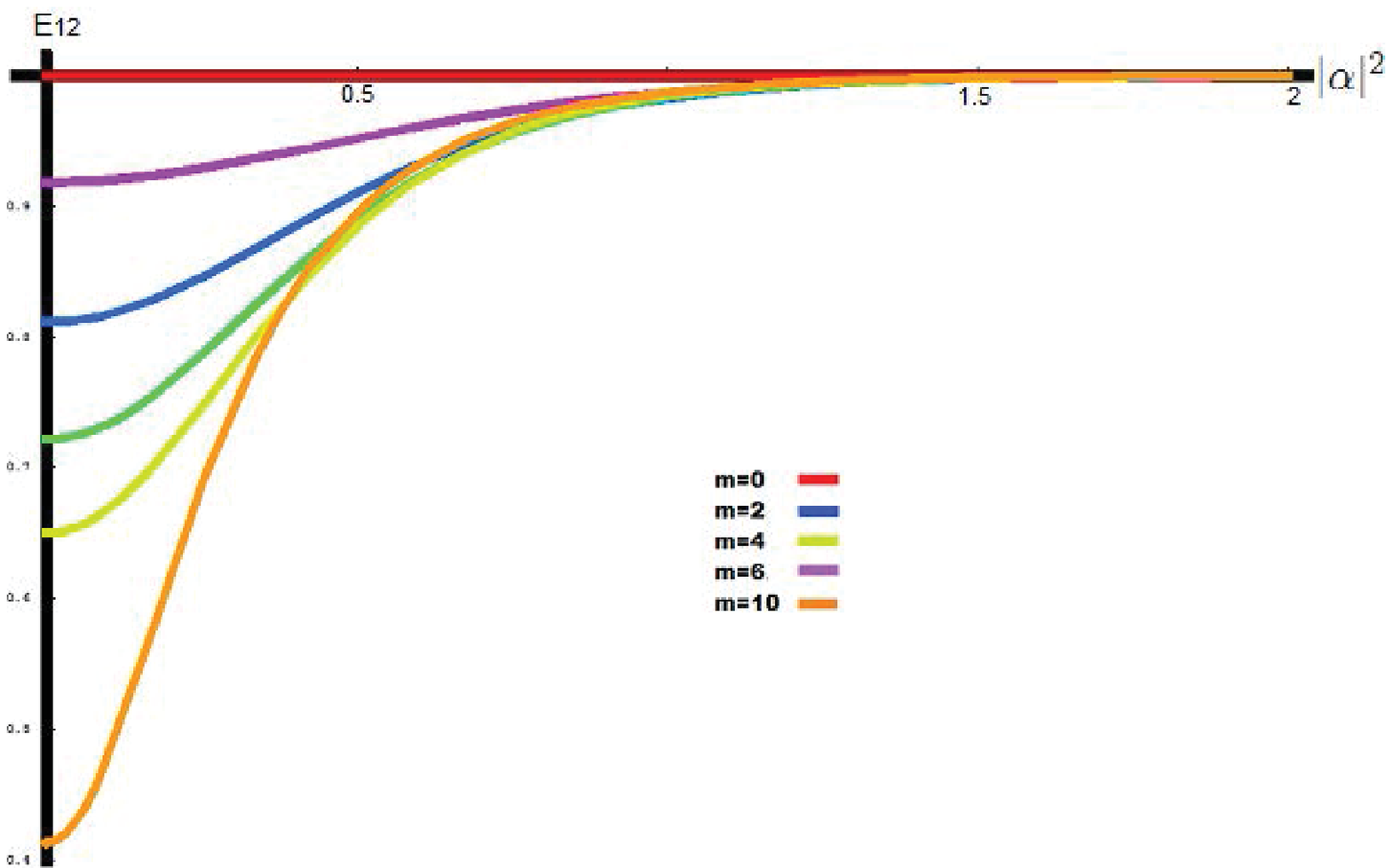}
\end{minipage}
\begin{minipage}[t]{3in}
\centering
 \includegraphics[width=3in]{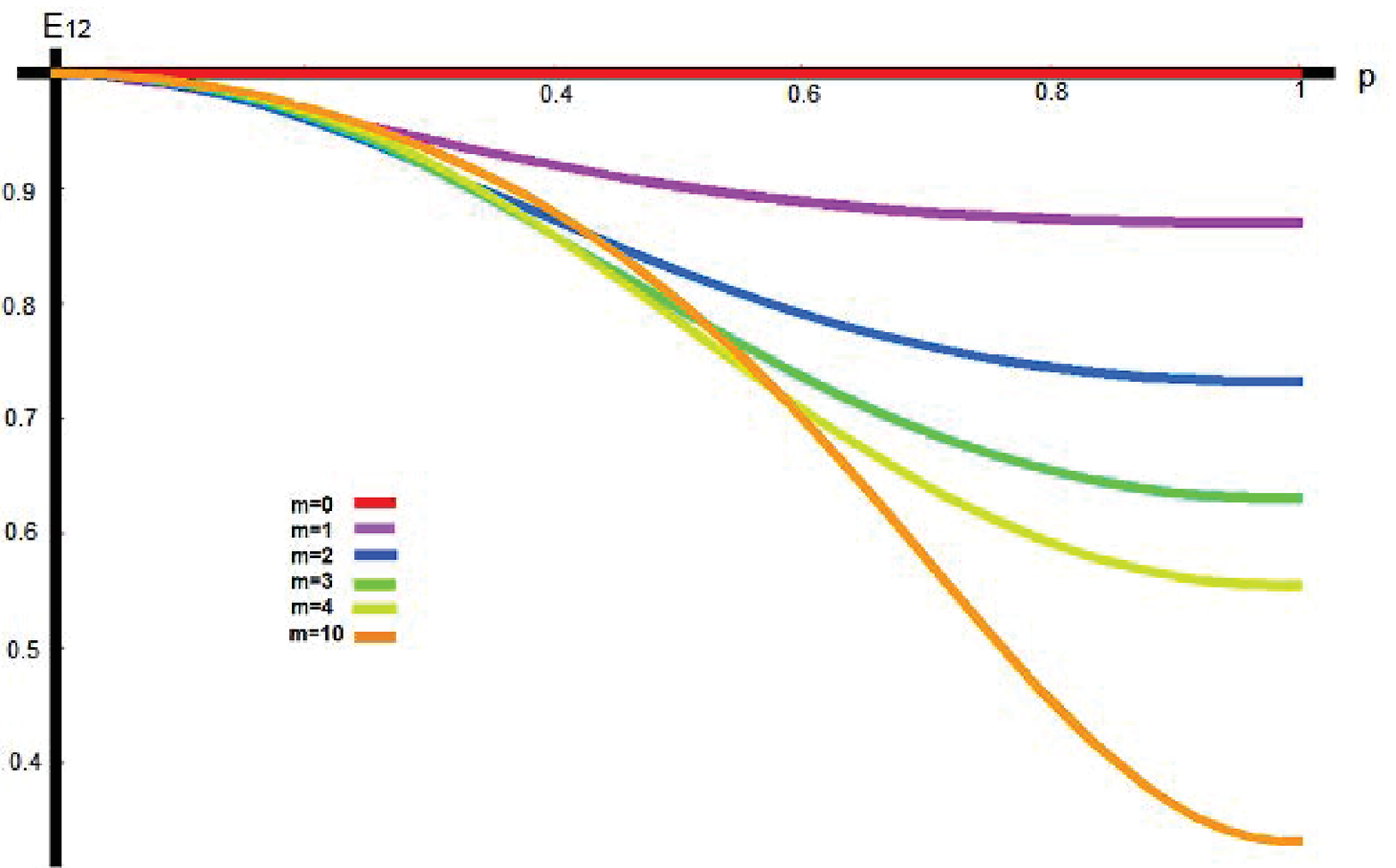}
\end{minipage}

{\bf Figure 2.}  {\sf The entanglement of formation $E_{12}$ versus
$|\alpha|^2$  and $p=e^{-2|\alpha|^2}$ for $k=1$ and different
values of photon excitation number $m$.}
\end{figure}

\section{Pairwise quantum correlations in excited quasi-GHZ coherent states}

\subsection{Photon added quasi-GHZ coherent states}
The analysis and results derived in the previous section are useful in
investigating pairwise quantum correlations in tripartite states involving
Glauber coherent states.  In this respect, we consider  the quasi-${\rm GHZ}$ coherent states  defined by
\begin{equation}\label{GHZ0}
|{\rm GHZ}_k(\alpha)\rangle=\mathcal{C}_{k} (\alpha) (|\alpha,\alpha,\alpha\rangle + e^{ik\pi} |-\alpha,-\alpha,-\alpha\rangle).
\end{equation}
where the normalization constant $\mathcal{C}_{k} $ is given by
\begin{equation}
\mathcal{C}^{-2}_{k} (\alpha) = 2 + 2 e^{-6|\alpha|^2} \cos k\pi.
\end{equation}
The excitation of the first mode by adding $m$ photon leads to the tripartite state
\begin{equation}
|| {\rm GHZ}_k(\alpha,m) \rangle= ( (a^+)^m
\otimes \mathbb{I} \otimes \mathbb{I})~ |{\rm GHZ}_k(\alpha)\rangle,
\end{equation}
from which we introduce the normalized photon added quasi-${\rm GHZ}$ coherent states as
\begin{equation}\label{GHZm0}
| {\rm GHZ}_k(\alpha,m) \rangle=  \frac{|| {\rm GHZ}_k(\alpha,m) \rangle}{\sqrt{\langle {\rm GHZ}_k(\alpha,m) || {\rm GHZ}_k(\alpha,m) \rangle}}.
\end{equation}
Using the expressions (\ref{valeur-moy}) and  (\ref{valeur-moy1}), the vector (\ref{GHZm0}) rewrites as
\begin{equation}\label{GHZm}
| {\rm GHZ}_k(\alpha,m) \rangle= \mathcal{C}_{k} (\alpha,m) (|m , \alpha \rangle \otimes |\alpha\rangle \otimes |\alpha\rangle,
 + e^{ik\pi} |m, -\alpha\rangle \otimes |-\alpha\rangle \otimes |-\alpha\rangle ) .
\end{equation}
where the normalization factor is
\begin{equation}
\mathcal{C}^{-2}_{k} (\alpha,m) = 2 + 2 \kappa_m e^{-6\vert \alpha \vert^2} \cos k\pi.
\end{equation}
For $m=0$, the state $| {\rm GHZ}_k(\alpha,m) \rangle$
(\ref{GHZm}) reduces to $| {\rm GHZ}_k(\alpha) \rangle$ (\ref{GHZ0}). It is also important to note
that for $|\alpha|$ large, the overlap between  Glauber coherent states $\vert \alpha \rangle$ and $\vert -\alpha \rangle$
approaches zero and then they are quasi-orthogonal. In this case, the state $| {\rm GHZ}_k(\alpha) \rangle$ (\ref{GHZ0})
reduces to the usual  {\rm GHZ} three qubit state
\begin{equation}\label{GHZusual}
| {\rm GHZ}_k(\infty) \rangle = \frac{1}{\sqrt{3}} (|{\bf 0} \rangle \otimes |{\bf 0}\rangle \otimes |{\bf 0}\rangle
 + e^{ik\pi} |{\bf 1} \rangle \otimes |{\bf 1} \rangle \otimes |{\bf 1} \rangle ),
\end{equation}
where $\vert {\bf 0} \rangle \equiv \vert \alpha \rangle $ and $\vert {\bf 1} \rangle \equiv \vert -\alpha \rangle$.

\noindent In investigating the pairwise quantum discord in a tripartite system $1-2-3$
prepared in the state $| {\rm GHZ}_k(\alpha,m) \rangle$, one needs the
reduced density matrices describing the two qubit subsystems $1-2$, $2-3$ and $1-3$. Since
only the first mode is affected by the photon excitations, it is simply seen that the reduced density
matrices $\rho_{12}={\rm Tr}_{3}\rho_{123}$ and $\rho_{13}={\rm Tr}_{2}\rho_{123}$ are
identical. The pure three mode density matrix  $\rho_{123}$ is given
\begin{equation}\label{matrixrho123}
\rho_{123} = | {\rm GHZ}_k(\alpha,m) \rangle \langle {\rm GHZ}_k(\alpha,m) |.
\end{equation}
After some algebra, the reduced density matrices $\rho_{12}$ and $\rho_{13}$ can be written as
\begin{eqnarray}\label{matrixrho12}
\rho_{12} = \rho_{13} = \frac{\mathcal{C}^2_{k}(\alpha,m)}{\mathcal{N}^2_{k}(\alpha,m)}\Bigg[ \bigg(\frac{1 + e^{-2|\alpha|^2}}{2}\bigg) |{\rm B}_{k}(\alpha,m)\rangle \langle {\rm B}_{k}(\alpha,m) |
+ \bigg(\frac{1 - e^{-2|\alpha|^2}}{2}\bigg) Z |{\rm B}_{k}(\alpha,m)\rangle \langle {\rm B}_{k}(\alpha,m) |Z \Bigg]
\end{eqnarray}
in terms of photon added quasi-Bell states (\ref{Bell-m}).  The operator $Z$ is the third Pauli generator defined by
$$Z |{\rm B}_{k}(\alpha,m)\rangle  = \mathcal{N}_{k}(\alpha,m)
[|m, \alpha ) \otimes|\alpha\rangle - e^{ik\pi} |m , -\alpha ) \otimes|-\alpha\rangle].$$
Similarly, by tracing out the first mode, the reduced matrix density $\rho_{23}$ takes the form
\begin{eqnarray}\label{matrixrho23}
 \rho_{23} = \frac{\mathcal{C}^2_{k}(\alpha,m)}{\mathcal{N}^2_{k}(\alpha,0)}\Bigg[ \bigg(\frac{1 + \kappa_m e^{-2|\alpha|^2}}{2}\bigg) |{\rm B}_{k}(\alpha,0)\rangle \langle {\rm B}_{k}(\alpha,0) |
+ \bigg(\frac{1 - \kappa_m  e^{-2|\alpha|^2}}{2}\bigg) Z |{\rm B}_{k}(\alpha,0)\rangle \langle {\rm B}_{k}(\alpha,0) |Z \Bigg].
\end{eqnarray}
To derive the pairwise correlation between the components of the
subsystems $1-2$, $2-3$ and $1-3$, we assume that the information is
encoded in even and odd Glauber coherent states (Schr\"odinger cat
states). In this sense, we introduce for the first mode the
following qubit mapping
\begin{eqnarray}\label{qubit1}
\vert  m , \pm \alpha \rangle= \sqrt{\frac{1 + \kappa_m e^{-2|\alpha|^2}}{2}} ~\vert 0 \rangle_1 \pm \sqrt{\frac{1 - \kappa_m e^{-2|\alpha|^2}}{2}} ~\vert 1 \rangle_1.
\end{eqnarray}
This coincides with the encoding scheme (\ref{qubit-m}) introduced in the previous section
to study the entanglement in quasi-Bell states. For the second and third modes, we consider the qubits defined by
\begin{eqnarray}\label{qubits23}
\vert  \pm \alpha \rangle = \sqrt{\frac{1 + e^{-2|\alpha|^2}}{2}} ~\vert 0 \rangle_i \pm \sqrt{\frac{1 - e^{-2|\alpha|^2}}{2}}~ \vert 1 \rangle_i, \qquad i = 2,3.
\end{eqnarray}
Substituting (\ref{qubit1}) and (\ref{qubits23}) in (\ref{matrixrho12}) (resp. (\ref{matrixrho23})), one can express the density matrix
$\rho_{12}$ (resp. $\rho_{23}$) in the two qubit basis $\{ \vert 0 \rangle_1 \otimes \vert 0 \rangle_2,
\vert 0 \rangle_1 \otimes \vert 1 \rangle_2, \vert 1 \rangle_1 \otimes \vert 0 \rangle_2 , \vert 1 \rangle_1 \otimes \vert 1 \rangle_2\}$
(resp.  $\{ \vert 0 \rangle_2 \otimes \vert 0 \rangle_3,
\vert 0 \rangle_2 \otimes \vert 1 \rangle_3, \vert 1 \rangle_2 \otimes \vert 0 \rangle_3 , \vert 1 \rangle_2 \otimes \vert 1 \rangle_3\}$).
The resulting density matrices have non-vanishing entries only along the diagonal and the anti-diagonal.

\subsection{Bipartite measures of quantum discord}

The state $| {\rm GHZ}_k(\alpha,m) \rangle$ (\ref{GHZm}) has rank
two reduced density matrices (\ref{matrixrho12})
 and (\ref{matrixrho23}). For these two qubit states, the Koashi-Winter relation which provides  the connection between the quantum discord and the entanglement of
formation,  can be exploited to obtain the relevant pairwise quantum
correlations. It is important to note that for two qubit states with
rank larger than two, the  derivation of quantum discord involves
optimization procedures that are in general complicated to achieve
analytically.

\noindent The total correlation in the subsystem  $1-2$ comprising
the optical modes 1 and 2,  is quantified by the mutual information
\begin{equation}\label{def: mutual information}
    I_{12}=S_{1}+S_{2}-S_{12},
\end{equation}
where $S_{12}$  is the von Neumann entropy of the quantum state
$\rho_{12}$ (\ref{matrixrho12}) ($S(\rho) = - {\rm Tr}~\rho \ln\rho$) and $S_{1}$ (resp. $S_{2}$) is the entropy
of the reduced state  $\rho_{1}={\rm Tr}_{2}( \rho_{12})$ (resp. $\rho_{2}={\rm Tr}_{1}( \rho_{12})$)  of the mode
$1$( resp. $2$). The mutual information $I_{12}$ contains both quantum and
classical correlations.  The classical correlations
$C_{12}$ can be determined by a local measurement optimization
procedure. To remove
the measurement dependence, a maximization over all possible
measurements is performed and the classical correlation writes
\begin{eqnarray}
    C_{12}  = S_2 - \widetilde{S}_{\rm min},
      \label{def: classical correlation}
\end{eqnarray}
where $\widetilde{S}_{\rm min}$  denotes the minimal value of the
conditional  entropy \cite{Henderson,Ollivier} (for more details, see the recent review \cite{Modi}). Thus, the quantum
discord,  defined as the difference between total correlation
$I_{12}$ and classical correlation $C_{12}$ \cite{Henderson,Ollivier},
 writes
\begin{equation} \label{discord-def-2}
    D_{12} = I_{12} - C_{12}
    =S_{1}+\widetilde{S}_{\rm min}-S_{12}.
\end{equation}
The main difficulty in deriving the analytical expression of
bipartite quantum discord (\ref{discord-def-2}), in  arbitrary mixed state, arises
in the minimization process of conditional entropy. This explains
why the explicit
expressions of quantum discord were obtained only for few
exceptional two-qubit quantum states, especially ones of rank two.
One may quote for instance the results obtained in
\cite{Ali,Adesso} (see also \cite{Rachid1,Rachid2,Rachid3}). Since the
density matrix $\rho_{12}$ (\ref{matrixrho12}) is of rank two,  the derivation  of the
analytical expression of
 $\widetilde{S}_{\rm min}$  in equation  (\ref{def: classical correlation}),  can be performed by purifying the
density matrix $\rho_{12}$ and  making use of Koashi-Winter relation
\cite{Koachi-Winter} (see also \cite{Shi1}). This relation
establishes the connection between the classical correlation of a
bipartite state $\rho_{12} $ and the entanglement of formation $E_{23}$ of
its complement $\rho_{23}$ in the pure state $\rho_{123}$ (\ref{matrixrho123}).  The minimal value of the conditional
entropy coincides with the entanglement of formation of $\rho_{23}$ \cite{Koachi-Winter}:
\begin{equation}\label{stild-min}
 \widetilde{S}_{\rm min} = E_{23}.
\end{equation}
The
Koaschi-Winter relation and the purification procedure provide us
with a computable expression of quantum discord in the bipartite state $\rho_{12}$
\begin{equation}\label{D12}
D_{12} =  S_1 - S_{12} + E_{23}
\end{equation}
when the measurement is performed on the subsystem $1$. The von Neumann entropy of the
reduced density $\rho_1 = {\rm Tr}_2 \rho_{12}$ is
\begin{equation}\label{S1}
S_1 = H\bigg(\frac{1}{2} \frac{(1 + \kappa_m  e^{-2|\alpha|^2})(1 + e^{-4|\alpha|^2}~\cos k\pi)}{1
+\kappa_m  e^{-6|\alpha|^2}\cos k\pi}\bigg),
\end{equation}
and the entropy of the bipartite density $\rho_{12}$ is explicitly
given by
\begin{equation}\label{S12}
S_{12} = H\bigg(\frac{1}{2} \frac{(1 + \kappa_m  e^{-4|\alpha|^2} \cos k\pi)(1 +
 e^{-2|\alpha|^2})}{1 +\kappa_m  e^{-6|\alpha|^2}\cos k\pi}\bigg).
\end{equation}
It is important to note that the entanglement of formation
measuring the entanglement of the subsystem $2$ with the ancillary
qubit, required in the purification process to minimize the
conditional entropy, is exactly the entanglement of formation
measuring the degree of intricacy between the optical modes $2$ and
 $3$. Using the definition of Wootters concurrence  (\ref{def-CONC}), one gets
\begin{equation}\label{C23}
C_{23}= \kappa_m  e^{-2|\alpha|^2}~\frac{(1- e^{-4|\alpha|^2})^2}{ 1 +
\kappa_m  e^{-6|\alpha|^2} \cos k \pi}
\end{equation}
and subsequently the corresponding entanglement of formation writes
\begin{equation}\label{E23}
E_{23}  = H\bigg(\frac{1}{2} + \frac{1}{2} \sqrt{ 1- \frac{\kappa^2_m  e^{-4|\alpha|^2}(1 -
e^{-4|\alpha|^2})^2}{(1 +\kappa_m  e^{-6|\alpha|^2}\cos k\pi)^2}}\bigg).
\end{equation}
Reporting (\ref{S1}), (\ref{S12}) and (\ref{E23}) in (\ref{D12}),
the quantum discord in the state $\rho_{12}$ is explicitly given by
\begin{eqnarray}\label{D12-explicit}
D_{12}&=&H\bigg(\frac{1}{2} \frac{(1 + \kappa_m  e^{-2|\alpha|^2})(1 + e^{-4|\alpha|^2}~\cos k\pi)}{1
+\kappa_m  e^{-6|\alpha|^2}\cos k\pi}\bigg)\\\nonumber
&-&H\bigg(\frac{1}{2} \frac{(1 + \kappa_m  e^{-4|\alpha|^2} \cos k\pi)(1 +
 e^{-2|\alpha|^2})}{1 +\kappa_m  e^{-6|\alpha|^2}\cos k\pi}\bigg)\\\nonumber
&+&H\bigg(\frac{1}{2} + \frac{1}{2} \sqrt{ 1- \frac{\kappa^2_m  e^{-4|\alpha|^2}(1 -
e^{-4|\alpha|^2})(1 - e^{-4|\alpha|^2})}{(1 +\kappa_m  e^{-6|\alpha|^2}\cos k\pi)^2}}\bigg),
\end{eqnarray}
The pairwise quantum discord existing in the mixed states $\rho_{23}$ (\ref{matrixrho23})
can be computed along  the same procedure. As result,  when  the measurement is performed on the
subsystem $2$,  the quantum discord is
\begin{equation}\label{D23}
 D_{23} =  S_2 - S_{23} + E_{13}.
\end{equation}
 The von Neumann entropy of the
reduced density $\rho_2 = {\rm Tr}_1 \rho_{12}$ is
\begin{equation}\label{S2}
S_2 = H\bigg(\frac{1}{2} \frac{(1 + e^{-2|\alpha|^2})(1 + \kappa_m  e^{-4|\alpha|^2}\cos k\pi)}{1
+\kappa_m  e^{-6|\alpha|^2}\cos k\pi}\bigg),
\end{equation}
and the entropy of the bipartite density $\rho_{23}$ is
\begin{equation}\label{S23}
S_{23} = H\bigg(\frac{1}{2} \frac{(1 + e^{-4|\alpha|^2}\cos k\pi)(1 +
\kappa_m  e^{-2|\alpha|^2} )}{1 +\kappa_m  e^{-6|\alpha|^2}\cos k\pi}\bigg).
\end{equation}
In purifying the state $\rho_{23}$ to derive the minimal amount
of conditional entropy, it is simple to show here also that the entanglement of formation
measuring the entanglement of the mode $3$ with an ancillary
qubit, is exactly the entanglement of formation
measuring the degree of intricacy between the optical modes  $1$ and  $3$. From (\ref{def-CONC}),  the concurrence between the modes 1 and 3
takes the following form
\begin{equation}\label{conc-mixte}
C_{13}= e^{-2|\alpha|^2}~\frac{\sqrt{(1 - \kappa^2_m  e^{-4|\alpha|^2})(1- e^{-4|\alpha|^2})}}{ 1 +
\kappa_m  e^{-6|\alpha|^2} \cos k \pi},
\end{equation}
from which one gets
\begin{equation}\label{Ejk}
E_{13}  = H\bigg(\frac{1}{2} + \frac{1}{2} \sqrt{ 1- \frac{e^{-4|\alpha|^2}(1 -
\kappa^2_m  e^{-4|\alpha|^2})(1 - e^{-4|\alpha|^2})}{(1 +\kappa_m  e^{-6|\alpha|^2}\cos k\pi)^2}}\bigg).
\end{equation}
Finally, the expression of quantum discord in  the state $\rho_{23}$ is
\begin{eqnarray}\label{D23-explicit}
D_{23}&=&H\bigg(\frac{1}{2} \frac{(1 + e^{-2|\alpha|^2})(1 + \kappa_m  e^{-4|\alpha|^2}\cos k\pi)}{1
+\kappa_m  e^{-6|\alpha|^2}\cos k\pi}\bigg)\\\nonumber
&-&H\bigg(\frac{1}{2} \frac{(1 + e^{-4|\alpha|^2}\cos k\pi)(1 +
\kappa_m  e^{-2|\alpha|^2} )}{1 +\kappa_m  e^{-6|\alpha|^2}\cos k\pi}\bigg)\\\nonumber
&+&H\bigg(\frac{1}{2} + \frac{1}{2} \sqrt{ 1- \frac{e^{-4|\alpha|^2}(1 -
\kappa^2_m  e^{-4|\alpha|^2})(1 - e^{-4|\alpha|^2})}{(1 +\kappa_m  e^{-6|\alpha|^2}\cos k\pi)^2}}\bigg).
\end{eqnarray}

\subsection{Some special cases}

In order to analyze the influence of the photon excitation number
$m$ on the bipartite quantum discord $D_{12}$ (\ref{D12-explicit}) and  $D_{23}$ (\ref{D23-explicit}),
we first give the figures 3 and 4 representing respectively $D_{12}$ and $D_{23}$ for
symmetric states ($k=0$).
\begin{figure}[!ht]
\centering
\begin{minipage}[t]{3in}
\centering
\includegraphics[width=3in]{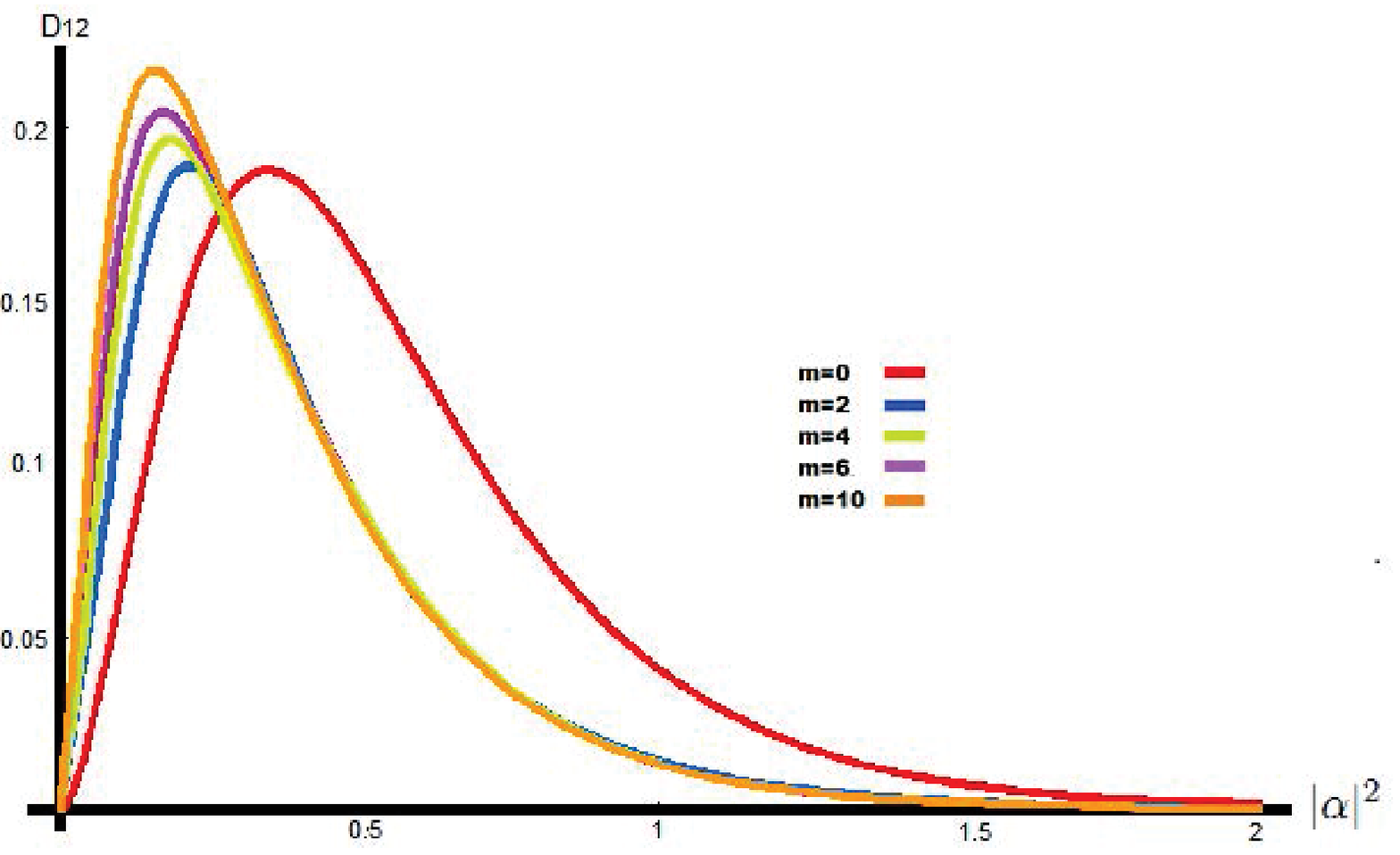}
\end{minipage}
\begin{minipage}[t]{3in}
\centering
 \includegraphics[width=3in]{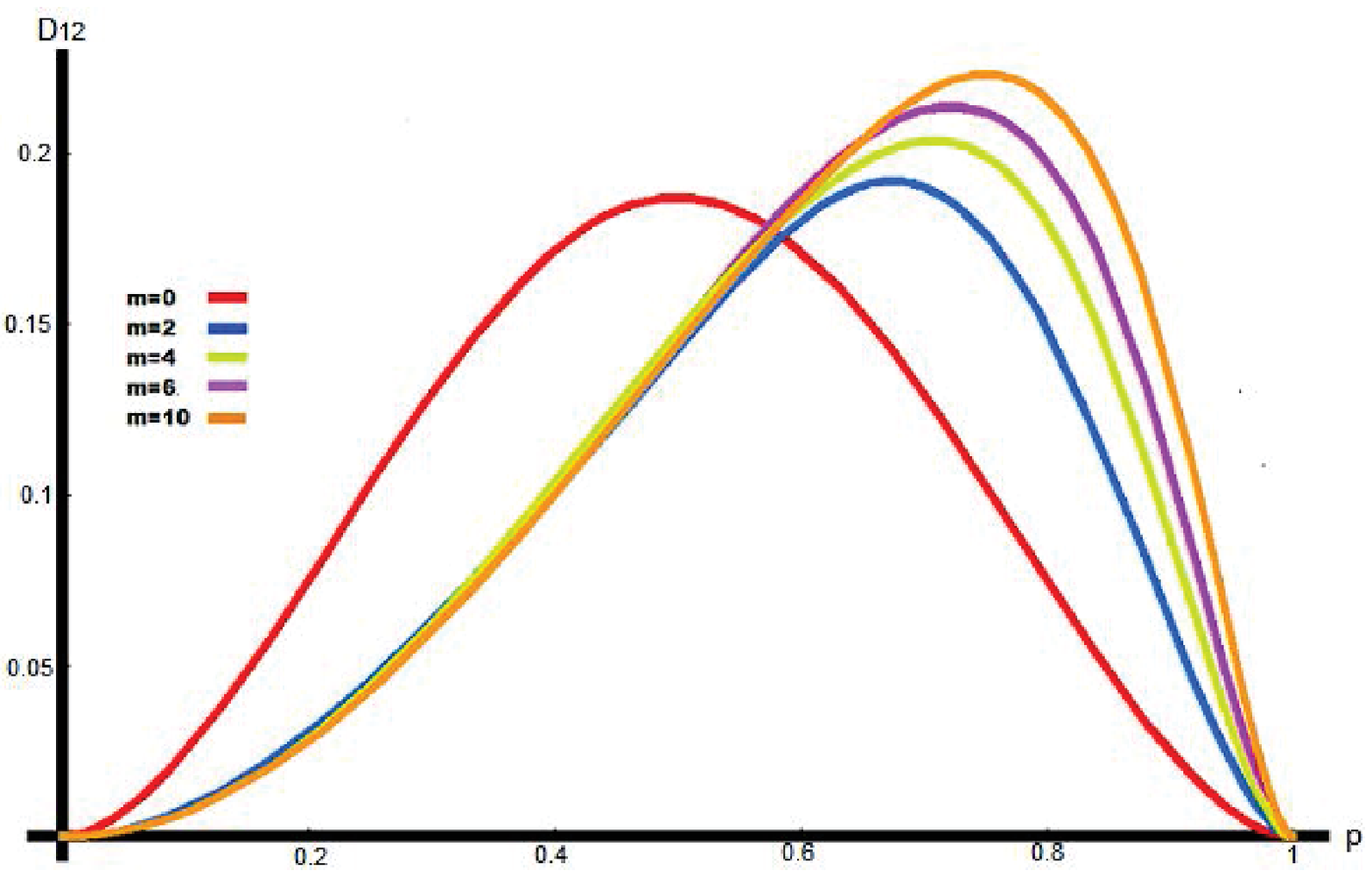}
\end{minipage}

{\bf Figure 3.}  {\sf The quantum discord $D_{12}$ versus
$|\alpha|^2$  and $p=e^{-2|\alpha|^2}$ for $k=0$ and different
values of photon excitation number $m$.}
\end{figure}
\begin{figure}[!ht]
\centering
\begin{minipage}[t]{3in}
\centering
\includegraphics[width=3in]{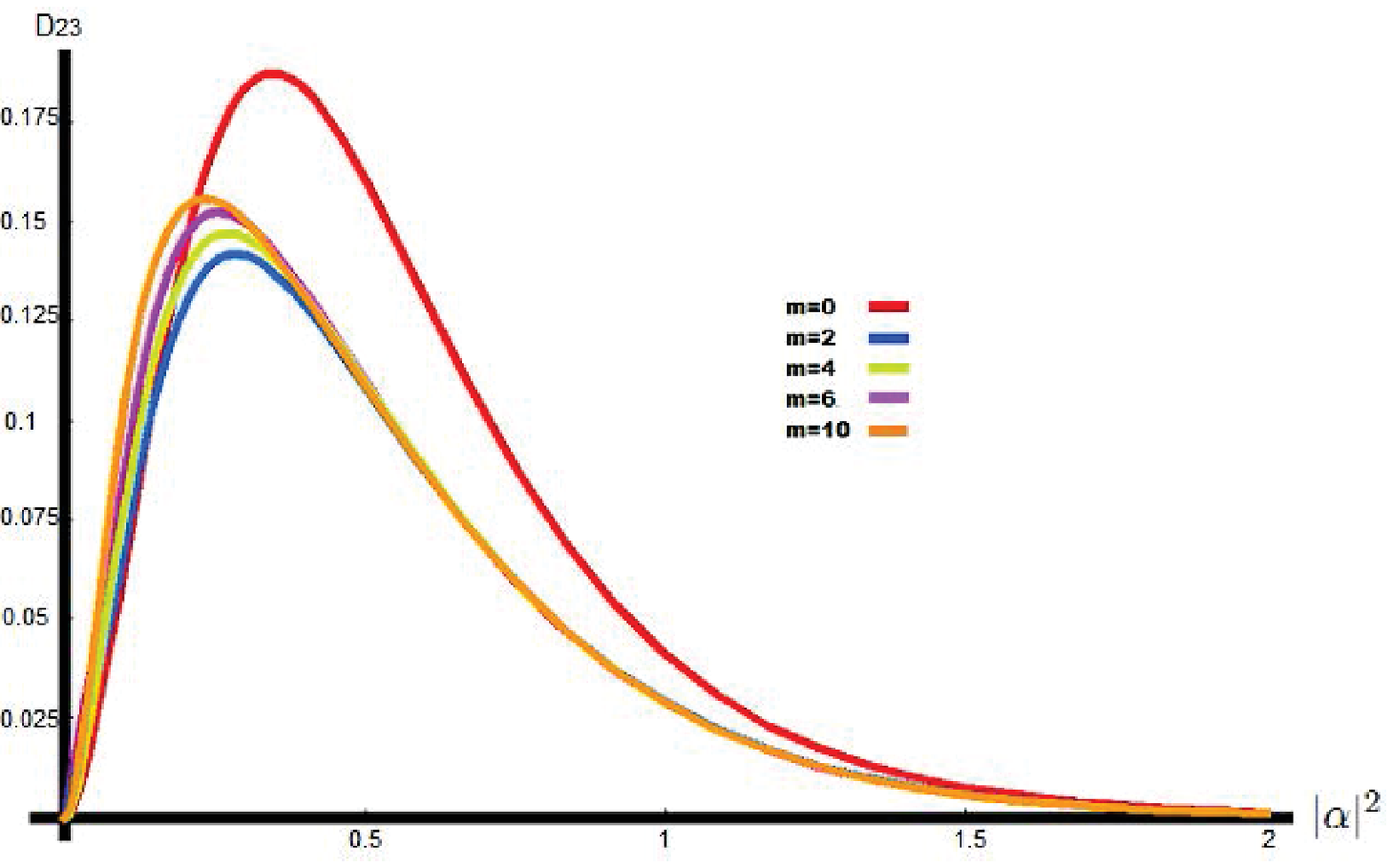}
\end{minipage}
\begin{minipage}[t]{3in}
\centering
 \includegraphics[width=3in]{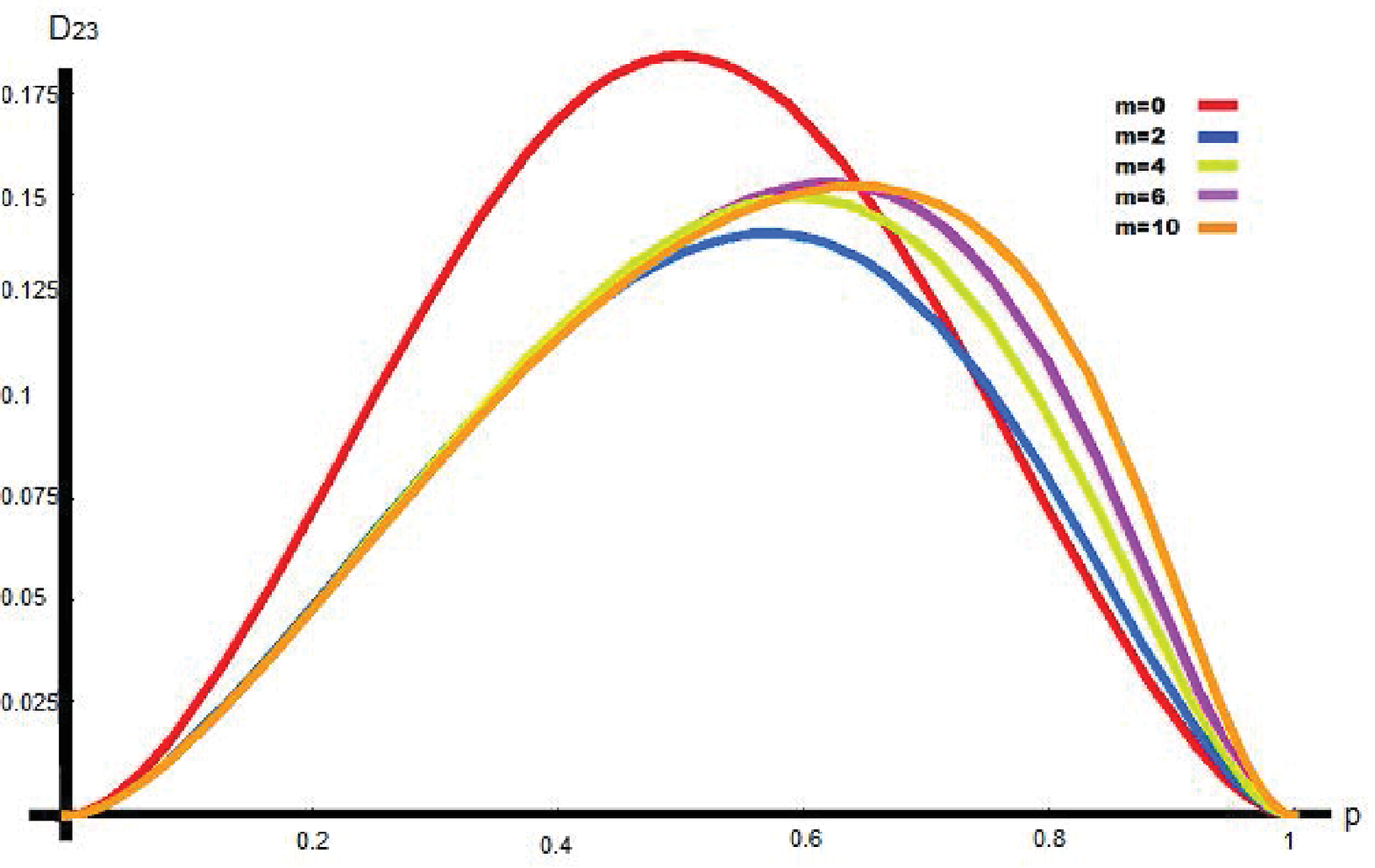}
\end{minipage}

{\bf Figure 4.}  {\sf The quantum discord $D_{23}$ versus
$|\alpha|^2$  and $p=e^{-2|\alpha|^2}$ for $k=0$ and different
values of photon excitation number $m$.}
\end{figure}

\noindent We can see from figure 3 that the quantum discord
$D_{12}(|\alpha|^2)$ between the optical modes 1 and 2, in the
symmetric case $(k=0)$, exhibits peaks which move to the left-hand
when the photon excitation number $m$ increases. It must be noticed
that the height of peaks, $D_{12}^{\rm max}(m)$,  increases with
increasing the number of added photons. We observe also that on the
left-hand side of the peak (weak regime), the quantum discord
$D_{12}$ rises rapidly with increasing the optical strength $\vert
\alpha \vert^2$. This indicates that the photon excitation of
Glauber coherent states, in the weak regime, induces an activation
of the correlations between the modes 1 and 2. In the strong regime
($\vert \alpha \vert$ large), the quantum discord tends  to zero
quickly as $m$ increases. The behavior of $D_{23}(|\alpha|^2)$ in
symmetric quasi-${\rm GHZ}$ coherent states $(k=0)$, depicted in
figure 4, shows that the maximal amount of quantum discord
$D_{23}^{\rm max}(m)$ is obtained for $m=0$ and $|\alpha|^2 \sim
0.5$. In contrast with $D_{12}$, $D_{23}^{\rm max}(m)$ decreases as
$m$ increases (figure 5). Thus, the increase of the quantum discord
$D_{12}$ is accompanied by a decrease of $D_{23}$ when the photon
excitation number $m$ increases. Remark also that for symmetric
states $(k=0)$, the photon excitation does not affect the amount of
pairwise quantum correlations $D_{12}$ and $D_{23}$ in the limiting
situations $|\alpha| \longrightarrow 0$ and $|\alpha|
\longrightarrow \infty$. This is no longer valid for antisymmetric
states $k=1$ especially for $|\alpha|$ approaching zero (see figures
5 and 6). Indeed, the quantum discord $D_{12}$ and $D_{23}$
decreases for $\alpha \longrightarrow  0$ as the ${\rm GHZ}$-like
coherent states become more  excited.

\begin{figure}[!ht]
\centering
\begin{minipage}[t]{3in}
\centering
\includegraphics[width=3in]{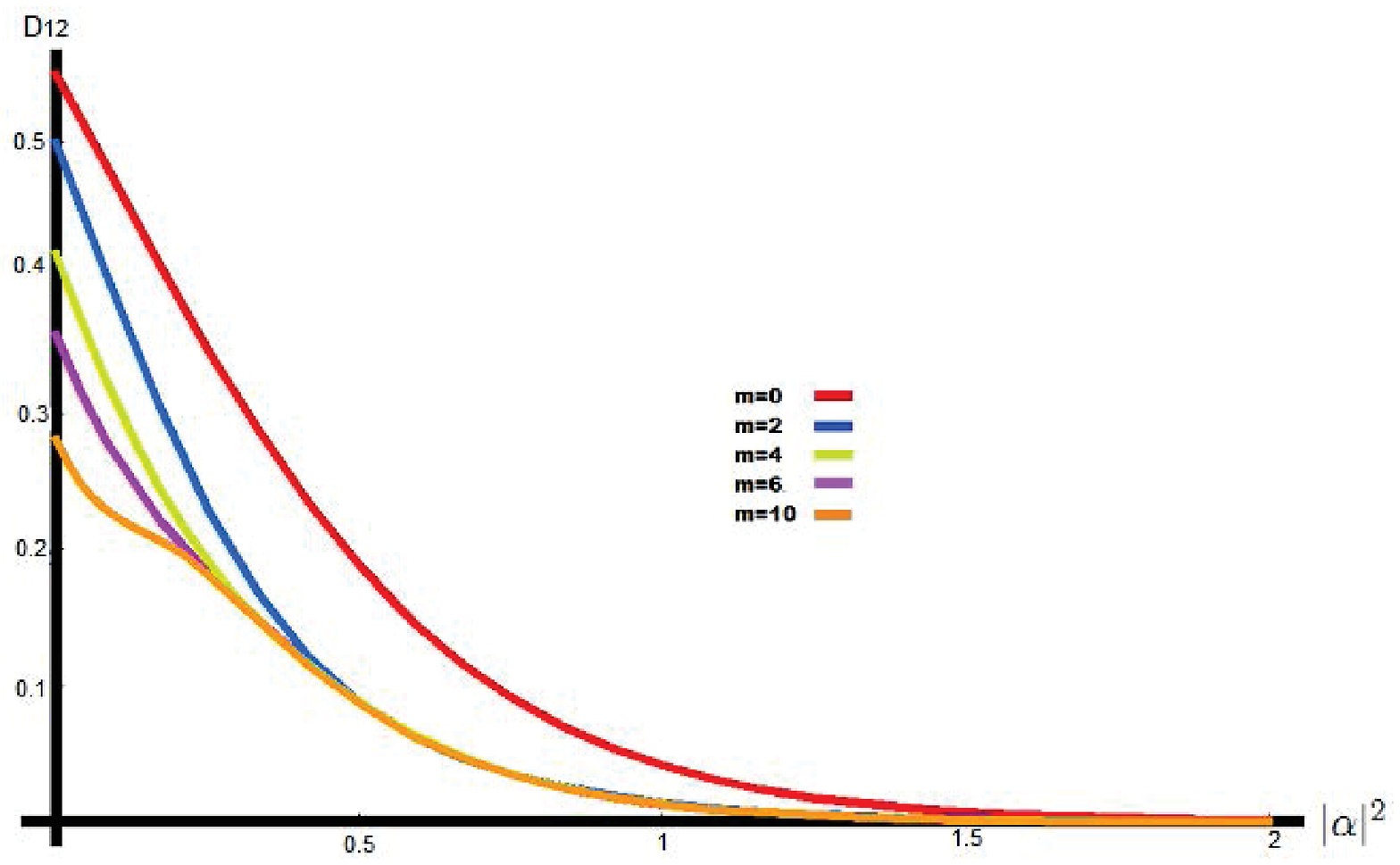}
\end{minipage}
\begin{minipage}[t]{3in}
\centering
 \includegraphics[width=3in]{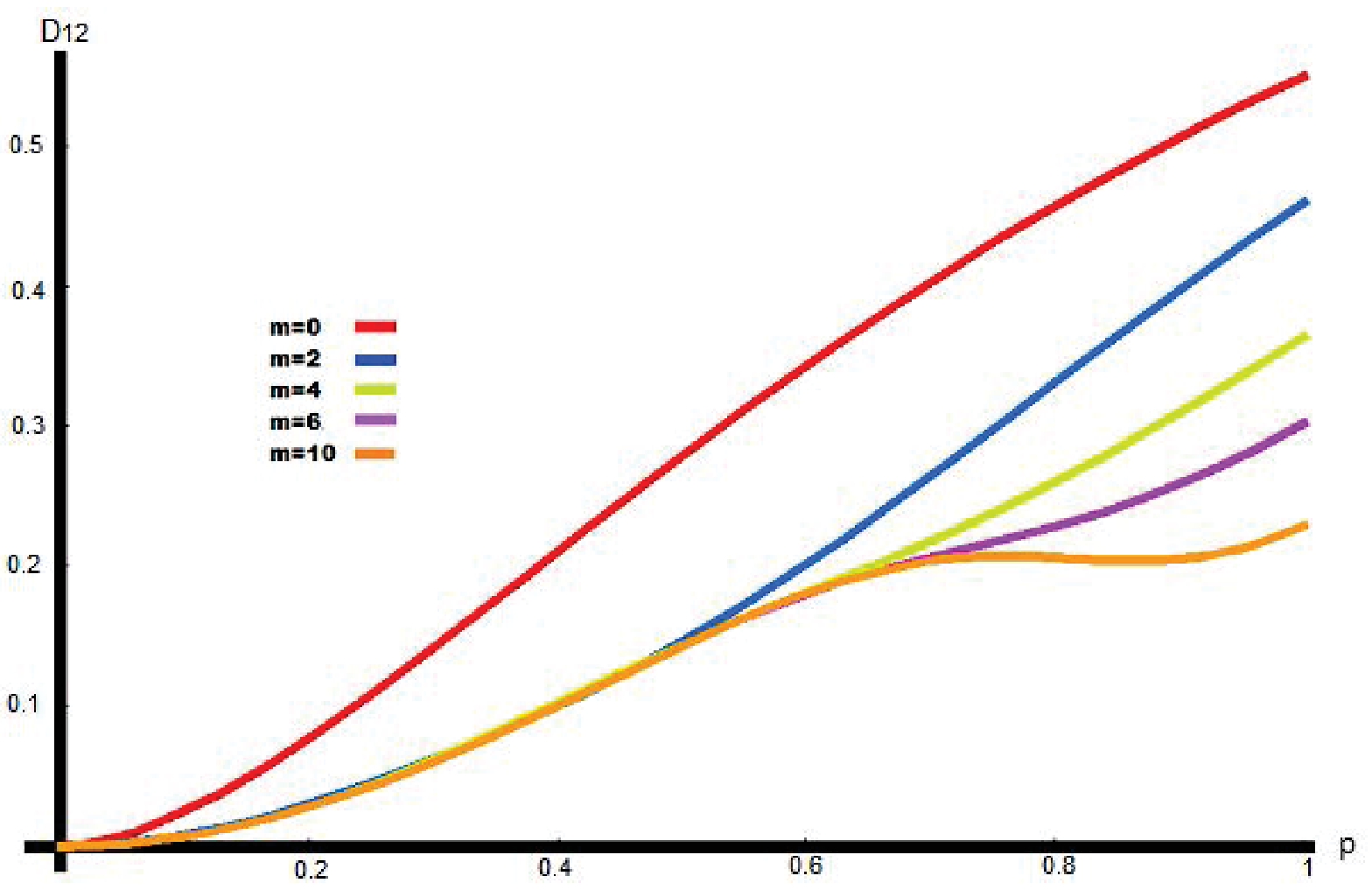}
\end{minipage}

{\bf Figure 5.}  {\sf The quantum discord $D_{12}$ versus
$|\alpha|^2$  and $p=e^{-2|\alpha|^2}$ for $k=1$ and different
values of photon excitation number $m$.}
\end{figure}

\begin{figure}[!ht]
\centering
\begin{minipage}[t]{3in}
\centering
\includegraphics[width=3in]{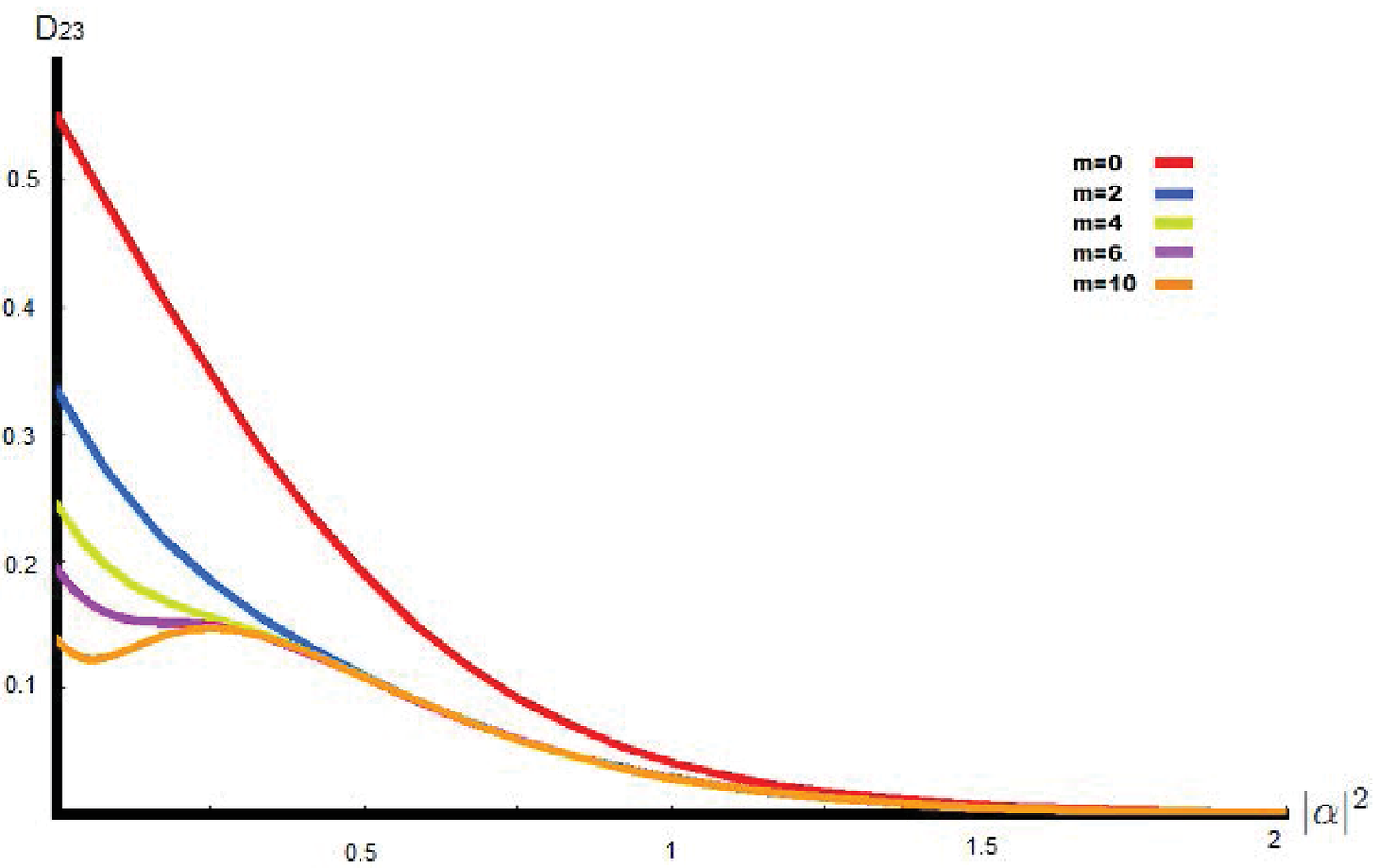}
\end{minipage}
\begin{minipage}[t]{3in}
\centering
 \includegraphics[width=3in]{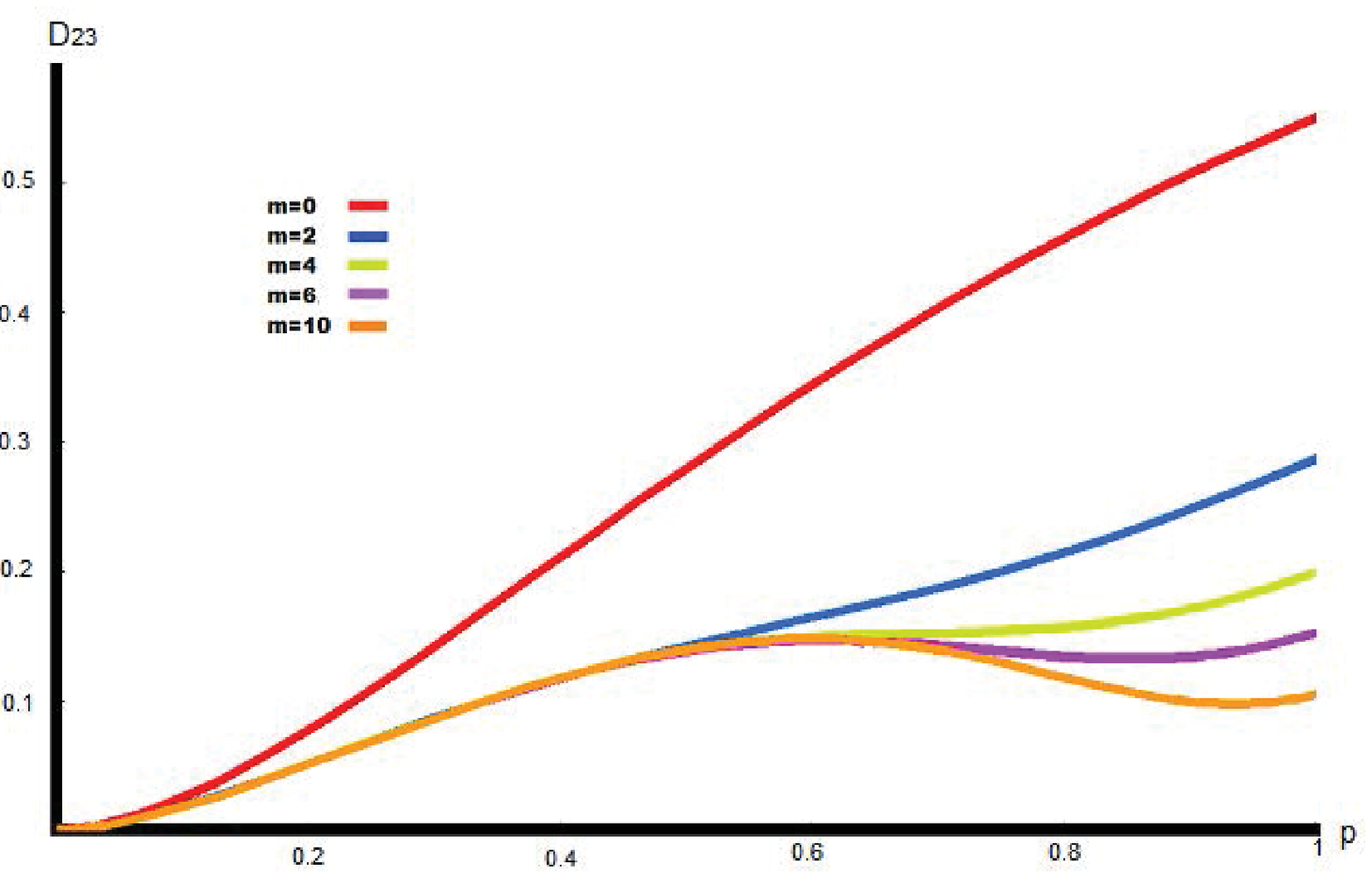}
\end{minipage}

{\bf Figure 6.}  {\sf The quantum discord $D_{23}$ versus
$|\alpha|^2$  and $p=e^{-2|\alpha|^2}$ for $k=1$ and different
values of photon excitation number $m$.}
\end{figure}

The behavior of quantum discord $D_{12}$ and $D_{23}$ in
anti-symmetric states $(k=1)$, when $|\alpha|$ approaches zero, can
be confirmed analytically. In fact, using (\ref{kappa-approx}), one
shows that for $|\alpha| \longrightarrow 0 $ the quantum discord
$D_{12}$ (\ref{D12-explicit}) and $D_{23}$ (\ref{D23-explicit}) are
given by
\begin{eqnarray}\label{D12-alpha=0}
D_{12}= D_{13} = H\bigg(\frac{2}{m+3}\bigg) - H\bigg(\frac{m+2}{m+3}\bigg) + H\bigg(\frac{1}{2} + \frac{1}{2}\frac{\sqrt{(m+1)(m+5)}}{m+3}\bigg),
\end{eqnarray}
and
\begin{eqnarray}\label{D12-alpha=0}
D_{23} = H\bigg(\frac{m+2}{m+3}\bigg) - H\bigg(\frac{2}{m+3}\bigg) + H\bigg(\frac{1}{2} + \frac{1}{2}\frac{\sqrt{m^2+2m+5}}{m+3}\bigg).
\end{eqnarray}
It is interesting to note that the antisymmetric photon added ${\rm GHZ}$-type coherent states $| {\rm GHZ}_1(\alpha,m) \rangle$ (\ref{GHZm})  reduces, for
$\vert \alpha \vert \longrightarrow 0 $, to
\begin{eqnarray}\label{GHZalpha=0}
| {\rm GHZ}_1(0,m) \rangle = \frac{1}{\sqrt{m+3}} (\sqrt{m+1} \vert m+1, 0, 0 \rangle + \vert m, 1, 0 \rangle + \vert m, 0, 1 \rangle )
\end{eqnarray}
which coincides with the usual three qubit ${\rm W}$ states for $m=0$ \cite{Dur}. The state (\ref{GHZalpha=0}) is expressed in the Fock-Hilbert basis. Hence,
according to the results plotted in figures 5 and 6, one concludes that photon excitations diminish the pairwise quantum
correlations existing  in three qubit states of ${\rm W}$ type.

\section{Monogamy of quantum discord in a three-qubit entangled state of ${\rm GHZ}$-type}

Having investigated the pairwise quantum discord in the state $|{\rm
GHZ}_k(\alpha,m)\rangle$ (\ref{GHZm}), we shall consider the
distribution of quantum correlations among its three optical modes.
It is well established that in a multi-qubit quantum system, the
monogamy property imposes restrictive constraints for the qubits to
share freely quantum correlations. Now, it is well established that,
unlike  the square of concurrence and the squashed entanglement, the
quantum discord does not follow the monogamy relation. In this
section, we investigate the influence of photon excitation number
$m$ on monogamy relation of quantum entropy-based quantum discord in
tripartite state of type $| {\rm GHZ}_k(\alpha,m) \rangle$
(\ref{GHZm}).

\noindent The entropy-based quantum discord, in the three modes
states ${\rm GHZ}_k(\alpha,m)$, is monogamous if, and only if, the
quantum discord deficit defined by
\begin{equation}\label{Delta123}
\Delta_{123}=\Delta_{123}(m,|\alpha|^{2})=D_{1\mid23}- D_{12} - D_{13},
\end{equation}
is positive. This condition reflects that
the monogamy property is satisfied when the quantum discord  $D_{1\mid23}$ between the first mode and the modes 2-3 (viewed as
a single subsystem) exceeds the sum of pairwise quantum discord  $D_{12}$  and $D_{13}$. We recall that
the concept of monogamy was originally introduced  by
Coffman, Kundu and Wootters in 2001 \cite{Coffman} in analyzing the
distribution of entanglement in a tripartite qubit system. After, several works
considered the monogamy of other quantum correlations quantifiers. In this section,
we shall determine the conditions under which the quantum discord satisfies the monogamy
property and a special attention will be devoted  to the influence of photon excitations
of Glauber coherent states. For this end, one has
to determine the pairwise quantum discord $D_{1\mid23}$ in the pure state $|{\rm GHZ}_k(\alpha,m)\rangle$ (\ref{GHZm}).
In  pure states  the quantum discord coincides
with the entanglement of formation. Hence, to compute  the entanglement between
qubit (1) and the joint qubits (23), we introduce the
orthogonal basis $\{ \vert 0 \rangle_1 , \vert 1 \rangle_1\}$
defined by
\begin{equation}\label{base1}
\vert 0 \rangle_1 = \frac{ \vert \alpha , m  \rangle +  \vert
-\alpha , m \rangle }{\sqrt{2(1 + \kappa_m e^{-2|\alpha|^2})}},
   \qquad \vert 1 \rangle_1 = \frac{\vert \alpha , m   \rangle  -  \vert -\alpha , m   \rangle }{{\sqrt{2(1-  \kappa_m e^{-|\alpha|^2})}}},
\end{equation}
for the first subsystem. For the  modes $(23)$, grouped into a single subsystem, we introduce the
orthogonal basis $\{ \vert  0 \rangle_{23} , \vert  1
\rangle_{23}\}$ given by
\begin{equation}\label{base2}
\vert 0 \rangle_{23} = \frac{ \vert \alpha, \alpha \rangle +  \vert -\alpha, -\alpha
\rangle}{\sqrt{2(1 + e^{-4|\alpha|^2})}}
   \qquad \vert 1 \rangle_{23} = \frac{ \vert \alpha, \alpha \rangle - \vert -\alpha, -\alpha
\rangle}{\sqrt{2(1 - e^{-4|\alpha|^2})}}.
\end{equation}
Inserting (\ref{base1}) and (\ref{base2}) in $\vert {\rm GHZ}_k(\alpha,m)\rangle$ , we
get the expression of the pure state $|{\rm GHZ}_k(\alpha,m)\rangle$  in the basis
$\{ \vert 0 \rangle_{1} \otimes \vert 0 \rangle_{23} ,
 \vert 0 \rangle_{1} \otimes \vert 1 \rangle_{23} , \vert  1 \rangle_{1}
 \otimes \vert 0 \rangle_{23} , \vert  1 \rangle_{1} \otimes \vert  1
 \rangle_{23}\}$. Explicitly, it is given by
\begin{equation}
\vert {\rm GHZ}_k(\alpha,m)
\rangle = \sum_{\alpha= 0,1} \sum_{\beta= 0,1} C_{\alpha,\beta}
\vert \alpha \rangle_1 \otimes \vert \beta
\rangle_{23}\label{mapping1}
\end{equation}
where the coefficients $C_{\alpha,\beta}$ are
$$ C_{0,0} = {\cal C}_k(\alpha , m)(1 + e^{ik\pi}) c^+_{1}c^+_{23}  , \qquad  C_{0,1} =  {\cal C}_k(\alpha , m) (1 -e^{ik\pi}) c^+_{1}c^-_{23}, $$
$$ C_{1,0} = {\cal C}_k(\alpha , m) (1 - e^{ik\pi}) c^+_{23}c^-_{1}  , \qquad  C_{1,1} =  {\cal C}_k(\alpha , m) (1 + e^{ik\pi}) c^-_{1}c^-_{23}, $$
in terms of the quantities
$$c^{\pm}_1 =\sqrt{\frac{1 \pm  \kappa_m e^{-2|\alpha|^2}}{2}}, \qquad  c^{\pm}_{23} =\sqrt{\frac{1 \pm e^{-4|\alpha|^2}}{2}}. $$
The concurrence between the two logical qubits $1$ and $23$ is given by
\begin{equation}\label {C123}
 C_{1\mid23} = \frac{\sqrt{(1- \kappa^2_m e^{-4|\alpha|^{2}})(1- e^{-8|\alpha|^{2}})}}{1+ \kappa_m e^{-6|\alpha|^{2}}\cos k\pi},
\end{equation}
from which we obtain
\begin{equation}\label{D123}
   D_{1\mid23} =E_{1\mid23}
   =H\bigg(\frac{1}{2}+\frac{1}{2}\frac{\kappa_m e^{-2|\alpha|^{2}} + e^{-4|\alpha|^{2}}\cos k\pi}{1+ \kappa_m e^{-6|\alpha|^{2}}\cos k\pi}\bigg).
\end{equation}
Inserting $D_{1\mid23}$ (\ref{D123}) and $D_{12} = D_{13}$
(\ref{D12}) in (\ref{Delta123}), one gets the explicit expression of
the  quantum discord deficit $\Delta_{123}$. The corresponding
behavior  as function of $|\alpha|^2$ (and $p=e^{-2\vert \alpha
\vert^2}$) for various values of photon excitation order $m$ is
displayed in the figures 7 and 8.

\begin{figure}[!ht]
\centering
\begin{minipage}[t]{3in}
\centering
\includegraphics[width=3in]{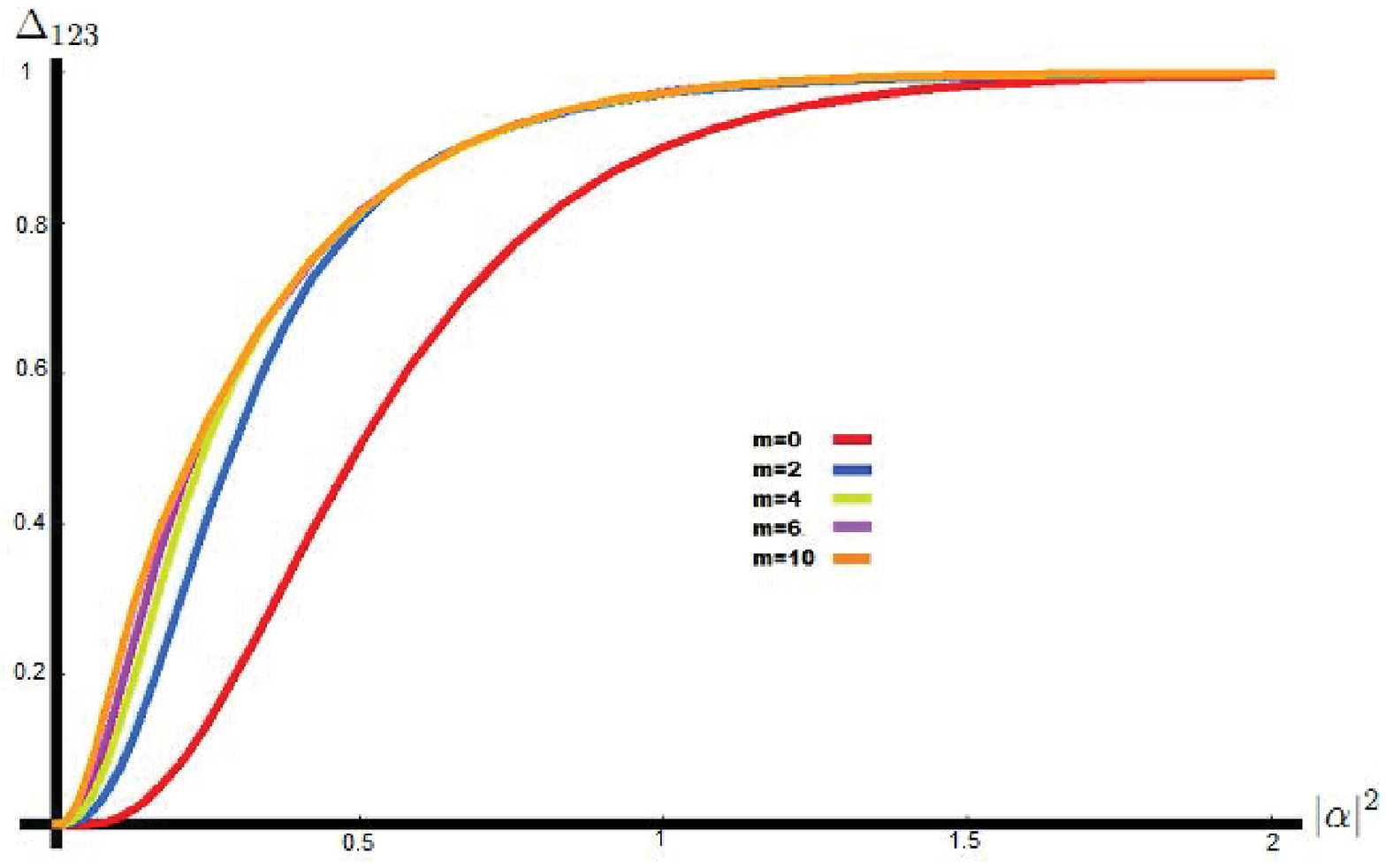}
\end{minipage}
\begin{minipage}[t]{3in}
\centering
 \includegraphics[width=3in]{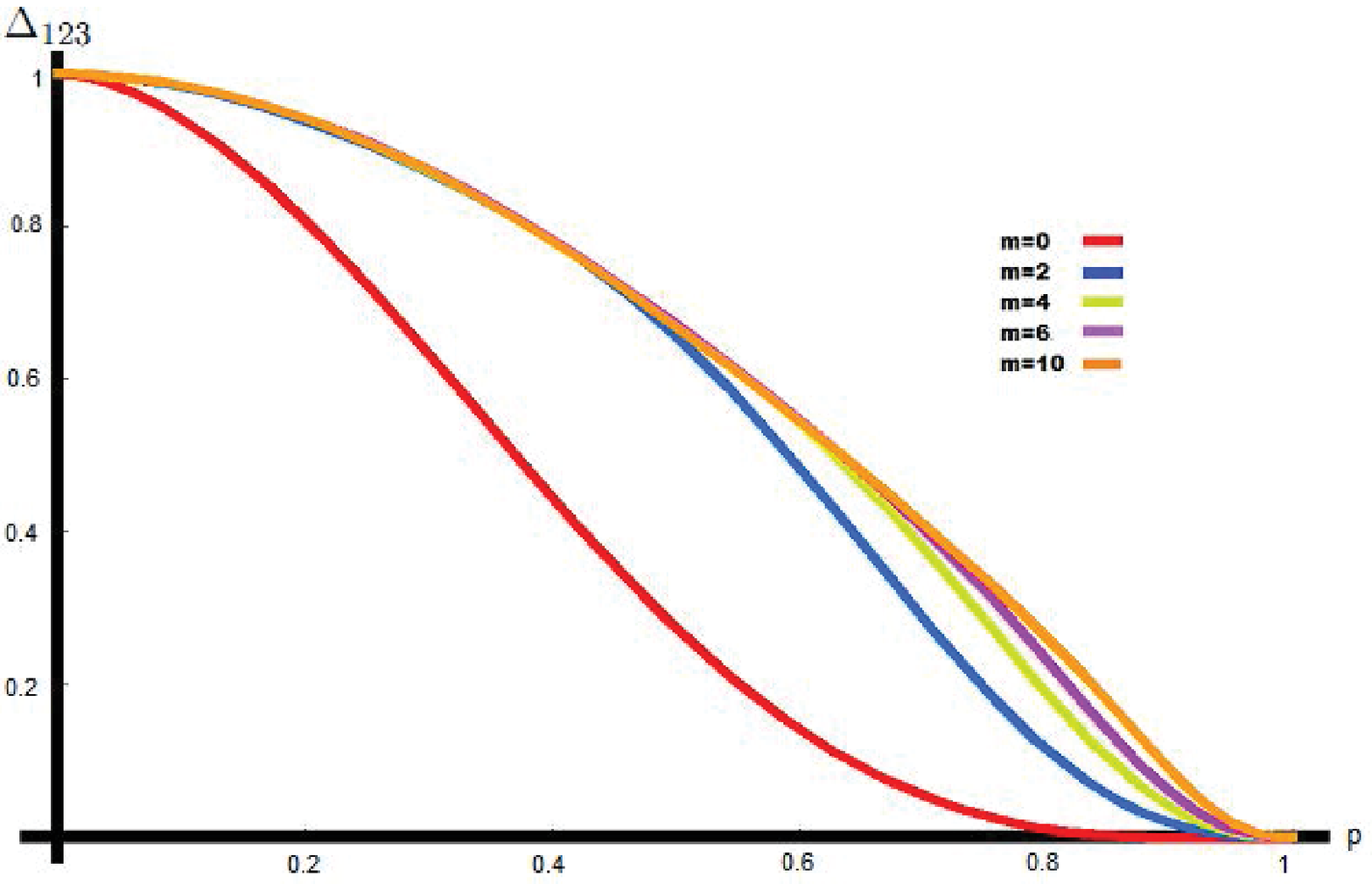}
\end{minipage}

{\bf Figure 7.}  {\sf The quantum discord deficit $\Delta_{123}$
versus $|\alpha|^2$  and $p=e^{-2|\alpha|^2}$ for $k=0$ and
different values of photon excitation number $m$.}
\end{figure}



\begin{figure}[!ht]
\centering
\begin{minipage}[t]{3in}
\centering
\includegraphics[width=3in]{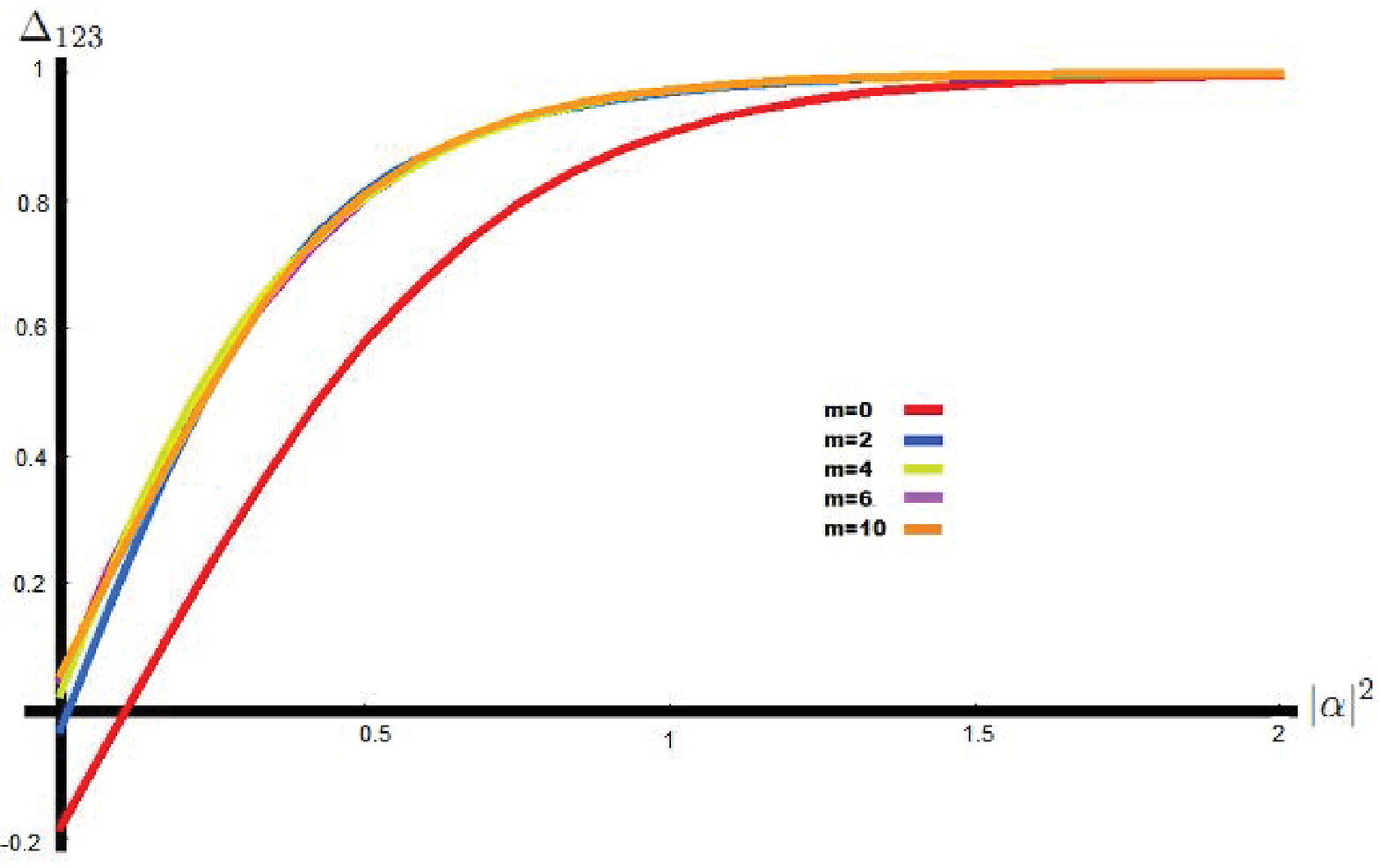}
\end{minipage}
\begin{minipage}[t]{3in}
\centering
 \includegraphics[width=3in]{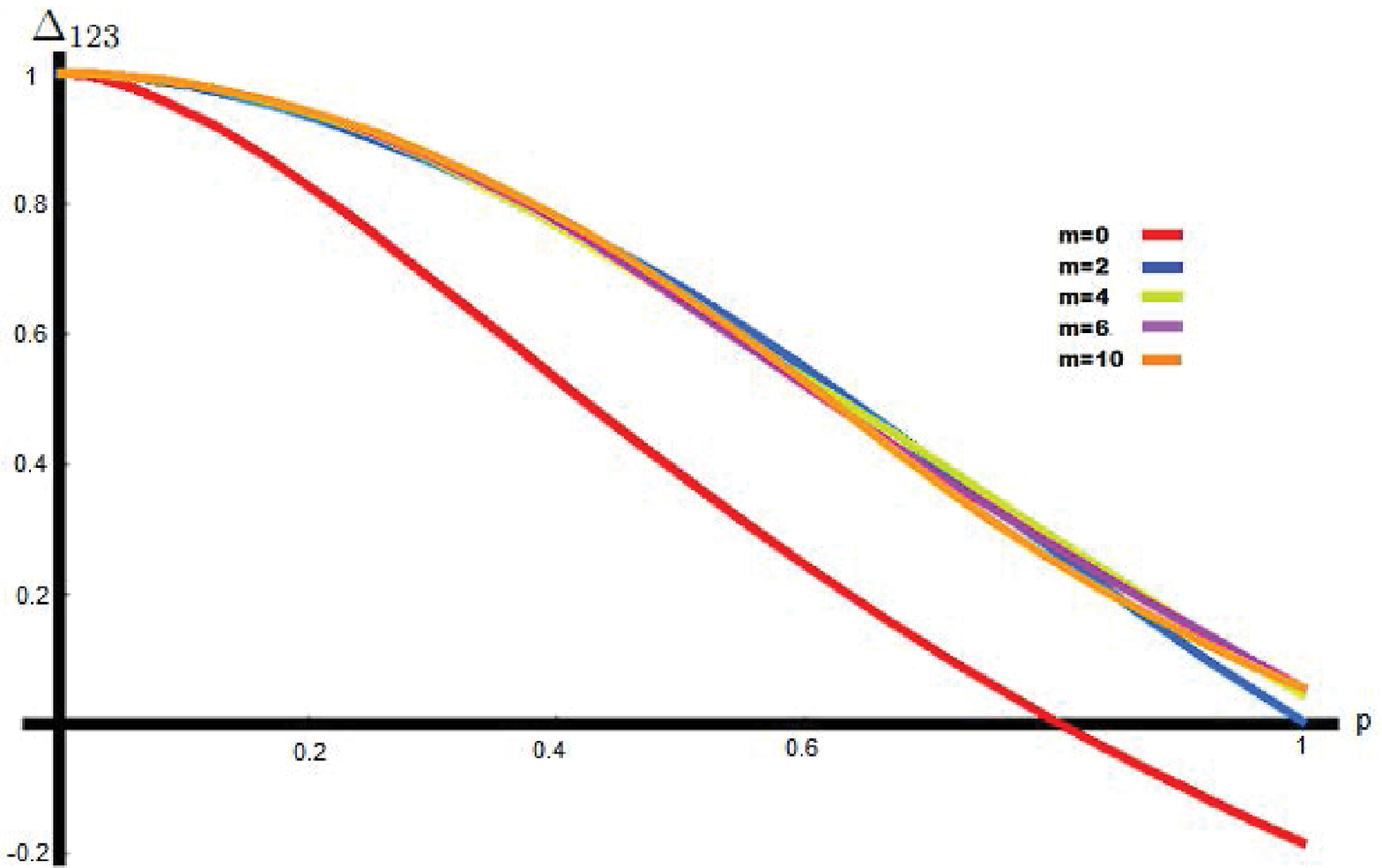}
\end{minipage}

{\bf Figure 8.}  {\sf The quantum discord deficit $\Delta_{123}$
versus  $|\alpha|^2$  and $p=e^{-2|\alpha|^2}$ for $k=1$ and
different values of photon excitation number $m$.}
\end{figure}


It can be inferred that the photon excitation of symmetric   quasi
{\rm GHZ}-coherent states $(k=0)$ does not affect the monogamy
property of quantum discord. The quantum discord deficit
$\Delta_{123} $  is always positive. The situation is slightly
different for antisymmetric quasi {\rm GHZ}-coherent states $(k=1)$.
In absence of photon excitation $(m=0)$, the quantum discord
violates the monogamy inequality for $|\alpha|^2 < 0.1075$ (or
equivalently $p > 0.806$). Remarkably, this violation tends to
disappear when photons are added and the quantum discord becomes
monogamous. For symmetric as well antisymmetric quasi {\rm
GHZ}-coherent states comprising $m\geq 2$ added photons,
$\Delta_{123}$  is almost identical in particular for high values of
$|\alpha|$ ($|\alpha|^2 \geq 1.5$). Remark that for $|\alpha|$ large
the photon added three mode coherent states $|{\rm
GHZ}_k(\alpha,m)\rangle$ tend to the usual
Greenberger-Horne-Zeilinger three qubit states (\ref{GHZusual}).
This indicates that, in this  case,
 photon addition process does not affect the distribution of the quantum correlations. Another special limiting situation
  concerns Glauber states with amplitude approaching zero.
For symmetric states $|{\rm GHZ}_0(\alpha = 0,m)$, it is easy to verify  from the  equations (\ref{D12}) and (\ref{D123}) that $\Delta_{123} = 0$ for any
photon excitation order $m$. For the antisymmetric
states $|{\rm GHZ}_1(\alpha = 0,m)$, which coincide with three qubit  states of ${\rm W}$-type (\ref{GHZalpha=0}),
the monogamy discord deficit increases  as $m$ increases (see figure 8). This result can be recovered analytically.
Indeed, for $k=1$ and $|\alpha| \longrightarrow 0$, one shows that
\begin{equation}\label{D123-alpha0}
   D_{1\mid23} \longrightarrow H\bigg(\frac{2}{m+3}\bigg),
\end{equation}
and using the result (\ref{D12-alpha=0}) one has
\begin{equation}\label{Delta123-alpha0}
   \Delta_{123} \longrightarrow  2 H\bigg(\frac{m+2}{m+3}\bigg) -2 H\bigg(\frac{1}{2} + \frac{1}{2}\frac{\sqrt{(m+1)(m+5)}}{m+3}\bigg) - H\bigg(\frac{2}{m+3}\bigg).
\end{equation}
The behavior $\Delta_{123}$ near the point $\alpha = 0$, plotted in the figure 8, reflects that the photon addition tends to increase the quantum deficit  $\Delta_{123}$
and subsequently to reduce the violation of monogamy relation  in states  of ${\rm W}$-type.


\section{ Concluding remarks}

In multipartite quantum systems, the monogamy is probably one of
the most important relation which imposes severe restriction on the
 structure of entanglement distributed
among many parties.   In this context, the main interest of this
paper was the monogamy property of quantum discord in three
qubit systems where the information is encoded in even and
odd Glauber coherent states. In particular, we investigated the influence of
 photon excitations  on the shareability of quantum discord between
the three optical modes of a quantum of ${\rm GHZ}$-type. We derived the
quantum discord deficit by evaluating analytically  the pairwise
correlations in terms of the photon excitation number and the
optical strength of Glauber coherent states. The symmetric quasi-${\rm GHZ}$ coherent
states follow the monogamy property for any photon excitation order.
We have also shown that the photon excitation of
antisymmetric quasi-${\rm GHZ}$ coherent
states reduces the violation of the monogamy property especially in states
involving Glauber coherent states with small amplitudes.\\

\noindent Finally, the investigation of the influence of photon
excitations on the monogamy of quantum correlations in the states of
{\rm GHZ}-type using geometric based quantifiers such as
Hilbert-Schmidt norm  or trace distance would be interesting. On the
other hand,  another significant issue which deserves to be examined
concerns the dynamics of quantum discord under  the effect of
subtracting photons on the pairwise correlations in multipartite
coherent states.

\end{document}